\documentstyle[prb,aps,epsf]{revtex}
\begin{document}
\draft
\preprint{Phys. Rev. B {\bf 52}, 1  Oct. 1995}
\tighten

%
%
\title{
	 Infrared conductivity in layered $d$-wave superconductors
}
\author{ M.\ J.\ Graf\cite{address}, Mario Palumbo and D.\ Rainer }
\address{Physikalisches Institut, Universit\"at Bayreuth,
	 D-95440 Bayreuth, Germany}
\author{ J.\ A.\ Sauls }
\address{Department of Physics and Astronomy, Northwestern University,
	 Evanston, Illinois 60208}
\maketitle
\begin{abstract}
We calculate the infrared conductivity of a stack of coupled,
two-dimensional superconducting planes within the Fermi liquid theory
of superconductivity. We include the effects of random scattering
processes and show that the presence of even a small
concentration of resonant impurities, in a $d$-wave superconductor, has
an important effect on both the in-plane and $c$-axis transport
properties, which could serve as signatures for $d$-wave pairing.
\end{abstract}
\pacs{\noindent PACS numbers:
74.20.Mn, 74.25.Nf, 74.72.-h, 74.80.Dm
\dotfill
Phys. Rev. B {\bf 52}, 1 {\sc October} 1995
}
\medskip
%
%
%
%
\section{INTRODUCTION} \label{Intro}
One of the central issues in the field of high temperature
superconductivity is the symmetry of the superconducting order
parameter. A number of recent experiments, specifically Josephson
interference experiments, suggest a $d$-wave order parameter with $d_{x^2-y^2}$
symmetry,\cite{wollman93,tsuei94,wollman94a,brawner94} while
experiments on $c$-axis tunneling\cite{sun94} are consistent with a
traditional $s$-wave order parameter. In this paper we calculate the
in-plane and $c$-axis conductivities for layered, $d$-wave
superconductors in the limit $q\to 0$ and for frequencies
$\hbar\omega\sim\Delta$ (infrared region). This long wavelength regime is
appropriate for superconductors with a mean free path that is short
compared to the penetration length, $\ell\ll\lambda$.  Given the small
coherence length in cuprates, the $q\to 0$ limit for the conductivity
easily accommodates the clean limit condition, $\xi\ll\ell$, which is
necessary for $d$-wave superconductivity not to be suppressed.
Calculations of the conductivity in the ultraclean limit
($\ell\gg\lambda\gg\xi$), which include polarization sensitivity to the
in-plane absorption and vertex corrections to the conductivity from
collective modes, will be discussed in a separate paper.

In order to represent the layered structure of high-$T_c$
superconductors, we consider the microscopic model of coupled
superconducting layers introduced in Ref.~\onlinecite{graf93}.  In this
model the charge carriers of each layer form a two-dimensional Fermi
liquid, which we describe by the quasiclassical theory of
superconductivity.  Thus, in-plane transport is dominated by charged
quasiparticles moving with an in-plane Fermi velocity, ${\bf v}_f$.
 Interlayer transport may originate from coherent
propagation,\cite{bulaevskii73,kats69,frick90,liu92,tanaka93} static and
thermally activated ({e.g.,} phonon assisted or pair fluctuation)
incoherent charge transfer,\cite{graf93,bulaevskii73,sugihara85,ioffe93}
resonant transfer,\cite{halbritter92} or several transfer mechanisms in
parallel.\cite{koshelev89,kumar90,kumar92,rojo94,rojo94b,varlamov94}
Interlayer transport has also been argued to take place through coherent pair
tunneling.\cite{chakravarty93} In this report we consider a model in
which the interlayer coupling is mediated through incoherent hopping
processes (the {\it interlayer diffusion model}), as in
Ref.~\onlinecite{graf93}.  These incoherent scattering events are
predominantly responsible for charge transfer in the $c$ direction and
Josephson coupling between the planes.  For simplicity, we consider a
circular Fermi surface for each conducting layer.
 This model is expected to be appropriate for
superconductors whose $c$-axis behavior is that of a stack of
superconductor-insulator-superconductor ($SIS$) Josephson junctions.
Such behavior has recently been observed experimentally in several
layered superconducting compounds.
\cite{kleiner92,kleiner94,kadowaki94,mueller94}

The infrared conductivity of three-dimensional bulk superconductors
with unconventional pairing has been calculated previously by Klemm
{\it et~al.},\cite{klemm88} Hirschfeld {\it et~al.},\cite{hirschfeld89}
and for two-dimensional superconductors by Hirschfeld, Putikka and
Scalapino \cite{hirschfeld94} using Green function methods;
our results agree with those found previously
whenever a direct comparison is possible. We calculate the infrared
conductivity for layered superconductors and find striking differences
between the conductivities of the $d$-wave and $s$-wave models. These
differences between $s$-wave and $d$-wave conductivities are not due
just to the differences in the density of states, which are known to
lead to different low-temperature behavior of the penetration depth,
NMR relaxation rates, etc.\cite{pines94} There are additional features
in the conductivity of $d$-wave superconductors which are attributed to
the formation of optically active Andreev bound states.

The outline of this paper is the following. In Sec.~\ref{Quasi_Th}, we
present our microscopic model in terms of the Fermi liquid theory of
superconductivity. Then in Sec.~\ref{Lin_Resp} we formulate the linear
current response to a spatially homogeneous, electromagnetic
perturbation. Section~\ref{Results} includes a discussion of
the phenomenological parameters of our model as well as the results of
the current response calculations. We relate the model parameters to
measurable normal-state quantities and calculate properties of the
superconducting state for both $s$-wave and $d$-wave pairing
interactions. We discuss the consistency between these two models and
experiments, and discuss the results of the linear response
calculations of the in-plane and $c$-axis infrared conductivities.

\section{QUASICLASSICAL THEORY} \label{Quasi_Th}

We perform the calculations within the quasiclassical theory of
superconductivity, first developed by Eilenberger,\cite{eilenberger68}
Larkin and Ovchinnikov,\cite{larkin68} and
Eliashberg.\cite{eliashberg71} This theory is capable of describing
both the equilibrium and dynamical properties of a Fermi liquid, in
either the normal or superconducting state. We briefly outline the
basic equations of this theory without discussing their justification;
readers interested in these details should see the original papers
mentioned above, or the recent review
articles.\cite{serene83,rammer86,larkin86,rainer92,muzikar94} We follow
closely the notation of Refs.~\onlinecite{graf93} and
\onlinecite{serene83}. In this report we are interested in the linear
response of a homogeneous, superconducting Fermi liquid to an electric
field described by a spatially uniform, time-dependent vector
potential. With this in mind, we start from the basic equations of the
quasiclassical theory neglecting the dependence on the space coordinate
(${\bf R}$).

\subsection{{Basic Equations}}

The most fundamental quantity in the quasiclassical theory of
superconductivity is the matrix propagator for quasiparticle excitations.
In order to describe nonequilibrium properties we use Keldysh's
formulation of dynamical phenomena (see Appendix).\cite{keldysh64}
This technique requires three types of propagators: retarded,
$\hat{g}^R$, advanced, $\hat{g}^A$, and Keldysh, $\hat{g}^K$.
These propagators are $2\times2$ matrices in particle-hole (Nambu) space,
\begin{equation}
{\hat g}^{R,A,K}\left(s;\epsilon,t\right) =
  \left(
        \begin{array}{cc}
          g^{R,A,K}       &  f^{R,A,K}       \\
          \bar{f}^{R,A,K} &  \bar{g}^{R,A,K}
        \end{array}
  \right) .
\label{g_nambu}
\end{equation}
{Note that we are limiting our considerations to distribution
functions which are spin scalars only; a more complete description of this
theory, including spin-dependent effects, can be found
elsewhere.\cite{serene83}}
For a spatially homogeneous Fermi liquid, the propagators
$\hat{g}^{R,A,K}\left(s;\epsilon,t\right)$ are functions of the Fermi surface
position, $s$, the single particle excitation energy, $\epsilon$, and
time, $t$. Their diagonal components are related to the spectrum and
distribution of Bogoliubov quasiparticle excitations, while their
off-diagonal components yield the Cooper-pair amplitude. In principle,
$\hat{g}^{K}$ contains all of the information about the relevant
measurable quantities (equilibrium and dynamic) in both the normal and
superconducting states. We make use of a compact notation which
combines the three Nambu matrices, $\hat{g}^{R,A,K}$, into a single
Nambu-Keldysh matrix,
\begin{equation}
{\hat g}\left(s;\epsilon,t\right) =
  \left(
        \begin{array}{cc}
          {\hat g}^{R}  &  {\hat g}^{K} \\
          0        &  {\hat g}^{A}
        \end{array}
\right) .
\label{g_nambu_keldysh}
\end{equation}
Thus, any operator written without superscript ({e.g.,} $\hat{a}$)
is interpreted as a Nambu-Keldysh matrix, while its retarded, advanced,
and Keldysh Nambu-matrix components are denoted by the appropriate
superscripts ({e.g.,} $\hat{a}^{R,A,K}$).

The central equation of the Fermi liquid theory of superconductivity is
the quasiclassical transport equation
\begin{eqnarray}
\displaystyle
\left[ \left(\epsilon + \frac{e}{c}{\bf v}_f\cdot{\bf
A}_{\ell}\right){\hat\tau}_3 -
       e\Phi_{\ell}\hat{1} - \hat\Delta_{\ell} -
\mbox{$\hat{\mbox{\footnotesize $\Sigma$}}$}_{\ell} \, , \,
       \hat{g}_{\ell}
\right]_\otimes = 0 ,
\label{keldysh_trans}
\end{eqnarray}
supplemented by the normalization condition
\begin{eqnarray}
\hat{g}_{\ell}\otimes\hat{g}_{\ell} = -\pi^{2}\hat{1} .
\label{g_norm_cond}
\end{eqnarray}
The $\hat\tau_3$ in equation~(\ref{keldysh_trans}) is to be understood as the
third Pauli matrix in particle-hole space combined with the unit matrix
in Keldysh space.
Equation (\ref{keldysh_trans}) generalizes the Landau-Boltzmann
transport equation for normal Fermi liquids to the superconducting
state. In our model we have a transport equation for each layer,
specified by the discrete index $\ell$. The spectrum and distribution
of quasiparticles in layer $\ell$ are affected by the in-plane vector
potential, ${\bf A}_\ell(t)$, the scalar potential, $\Phi_\ell(t)$, the
pairing field, $\hat\Delta_\ell(s,t)$, and the scattering self-energy,
$\mbox{$\hat{\mbox{\footnotesize $\Sigma$}}$}_\ell(s;\epsilon,t)$. The coupling
of quasiparticles (of charge
$e$) to an electromagnetic field appears explicitly in
equation~(\ref{keldysh_trans}), and also implicitly through the field
dependence of the self-energy, $\mbox{$\hat{\mbox{\footnotesize
$\Sigma$}}$}_\ell$. The Fermi velocity,
${\bf v}_f$, is a two-dimensional vector parallel to the planes. The
commutator in equation~(\ref{keldysh_trans}) is defined as
$[\hat{a},\hat{b}]_\otimes=\hat{a}\otimes\hat{b}-\hat{b}\otimes\hat{a}$,
where the symbol $\otimes$ denotes a folding product in the energy-time
domain (see Appendix), together with matrix multiplication of Nambu-Keldysh
matrices.  For details of this formalism see
Refs.~\onlinecite{serene83,rammer86,larkin86,rainer92,muzikar94}.

In the weak-coupling approximation, the retarded, advanced, and Keldysh
components of the pairing field take the form
\begin{eqnarray}
& &\hat{\Delta}_{\ell}^{R,A}\left(s;t\right) =
     \int\frac{d\epsilon}{4\pi i} \left< V\left(s,s'\right)
     \hat{f}^{K}_\ell\left(s';\epsilon,t\right) \right>_{s'} , \label{pair_SE}
\\
& &\hat{\Delta}_{\ell}^{K}\left(s;t\right) = 0 .
\end{eqnarray}
Here $V(s,s')$ is the pairing interaction, which determines both $T_c$ and
the symmetry of the order parameter, and $\hat{f}^{K}$ denotes the
off-diagonal part of the Keldysh propagator, $\hat{g}^{K}$.
The notation $\left<F(s)\right>_s$ denotes a Fermi surface average,
$\int d^2s\,n(s)F(s)$, where $n(s)$ describes the local density of
quasiparticle states on the Fermi surface, normalized so that
$\int d^2s\,n(s)=1$.
For the case of isotropic $s$-wave pairing, the pairing interaction takes
the form $V\left(s,s'\right)=V_{0}$, while for a $d$-wave model,
$V\left(s,s'\right)$ favors a pairing field with $d_{x^2-y^2}$ symmetry,
although essentially all of the results presented here are valid for any
unconventional order parameter with line nodes on the two-dimensional Fermi
surface.

The scattering self-energy of layer $\ell$, $\mbox{$\hat{\mbox{\footnotesize
$\Sigma$}}$}_\ell=
\mbox{$\hat{\mbox{\footnotesize
$\Sigma$}}$}^\|_\ell+\mbox{$\hat{\mbox{\footnotesize
$\Sigma$}}$}^\bot_{\ell,\ell-1}+\mbox{$\hat{\mbox{\footnotesize
$\Sigma$}}$}^\bot_{\ell,\ell+1}$,
describes both in-plane scattering ($\mbox{$\hat{\mbox{\footnotesize
$\Sigma$}}$}^\|_\ell$) and interplane
scattering ($\mbox{$\hat{\mbox{\footnotesize
$\Sigma$}}$}^\bot_{\ell,\ell\pm1}$). In-plane scattering
processes are taken into account via a self-energy of the form
\begin{equation}
\mbox{$\hat{\mbox{\footnotesize
$\Sigma$}}$}^{\|}_{\ell}\left(s;\epsilon,t\right) =
  c_i\,\hat{t}_\ell\left(s,s;\epsilon,t\right) ,
\label{sigma_scatt}
\end{equation}
where $c_i$ is an effective concentration of scattering centers.
We consider isotropic in-plane scattering, in which case
$\hat{u}_0=u_0\,\hat{1}$ is an isotropic scattering potential. The
$s$-dependence drops out of the in-plane $\hat t$ matrix so that
\begin{equation}
\hat{t}_\ell\left(\epsilon,t\right) = \hat{u}_{0} + \hat{u}_{0} \otimes
\left[N_{f}\left<
      \hat{g}_\ell\left(s;\epsilon,t\right)\right>_s\right] \otimes
  \hat{t}_\ell\left(\epsilon,t\right) ,
\label{t_matrix}
\end{equation}
where $N_f$ is the density of states at the Fermi energy.  The
scattering self-energy, $\mbox{$\hat{\mbox{\footnotesize
$\Sigma$}}$}^{\|}_{\ell}$, then contains only two
independent parameters. Following Buchholtz and
Zwicknagl,\cite{buchholtz81} we choose these two parameters to be the
normal-state scattering rate,
\begin{equation}
\frac{\hbar}{\tau_{\|}} = \frac{2\pi c_i N_{f}u_{0}^{2}}
                      {1 + \left(N_{f}u_{0}\pi\right)^{2}} ,
\label{tau_pllel}
\end{equation}
and the normalized scattering cross section,
\begin{equation}
\bar\sigma = \frac{\left(N_{f}u_{0}\pi\right)^{2}}
              {1 + \left(N_{f}u_{0}\pi\right)^{2}}\,,
\label{sigma_bar}
\end{equation}
which is the true cross section divided by the
cross section in the unitarity limit, $\sigma^{(u)}_{2d}=4/k_{f}$.
Thus, the normalized cross section ranges from $\bar\sigma=0$ for weak
scattering (Born limit), to $\bar\sigma=1$ for strong scattering
(unitarity limit). We have formulated the in-plane scattering
self-energy in terms of a $\hat{t}$ matrix. In this formulation, the
self-energy covers impurity scattering, including resonant scattering,
as well as any other type of scattering process, elastic or
inelastic, as long as it can be approximated by an effective lifetime
and an effective cross section. For instance, electron-phonon
scattering in this approximation corresponds to $\bar\sigma=0$
(Migdal's theorem) and a temperature dependent lifetime,
$\tau_{||}(T)$.

\begin{figure}
\begin{minipage}{0.8\hsize}
\epsfxsize=\hsize
\epsfbox{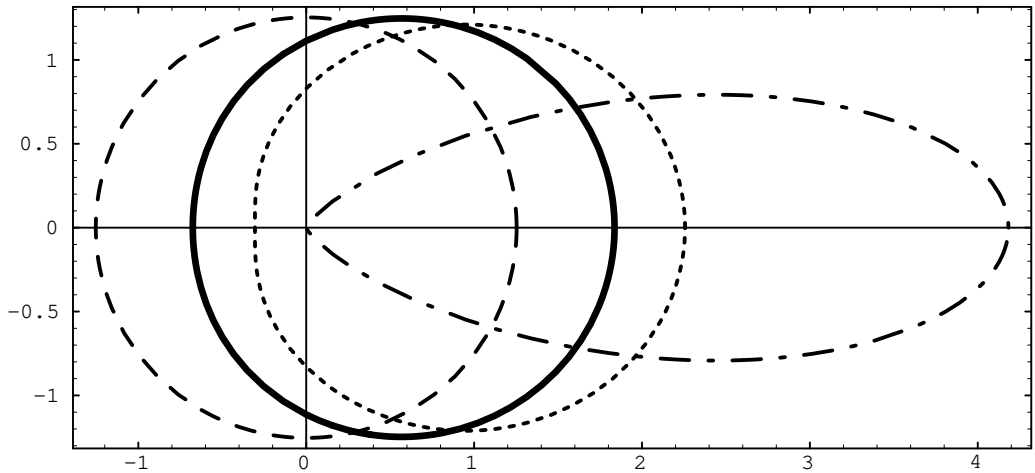}
\caption{
  Polar plot of the anisotropic scattering rate, $\tau_\bot^{-1}(s,s')$, which
is normalized here by the area for easier visualization.
  A quasiparticle traveling along the $x$ axis in layer $\ell$ (left to right)
 is scattered in a new direction in layer $\ell\pm1$ with a probability
proportional to the radial distance on this curve.
  The anisotropy parameter values, $\gamma$, shown are: 0 (dashed line),
0.5 (thick line), 1 (dotted line), 10 (dash-dotted line).}
\label{polar_plot}
\end{minipage}
\end{figure}

Since we neglect coherent transport along the $c$-axis (i.e., we set the
Fermi velocity along the $c$-axis to zero), the interlayer scattering
self-energies, $\mbox{$\hat{\mbox{\footnotesize
$\Sigma$}}$}^{\bot}_{\ell,\ell\pm1}$, are the only source of
interlayer coupling in the model.  We assume this coupling to be weak
(nonresonant), and thus describe interlayer scattering by a self-energy
in the Born approximation,
\begin{eqnarray}\label{hop}
\mbox{$\hat{\mbox{\footnotesize
$\Sigma$}}$}^{\bot}_{\ell,\ell\pm1}(s;\epsilon,t) &=&
 {\hat U}_{\ell,\ell\pm1}(t)\otimes
 \left[\frac{\hbar}{2\pi}\left\langle\frac{1}{\tau_\bot(s,s')}\,
     {\hat g}_{\ell\pm1}(s';\epsilon,t)\right\rangle_{\!\!\!s'} \right] \otimes
 {\hat U}^\dagger_{\ell,\ell\pm1}(t)
\ .
\end{eqnarray}
The interlayer self-energy is parametrized by an effective lifetime,
$\tau_\bot$, which is the characteristic time between consecutive
scattering processes between layers.\cite{graf94} The gauge operators,
${\hat U}_{\ell,\ell\pm1}$, and interlayer vector potentials,
$A^z_{\ell,\ell\pm1}$, are defined by
\begin{eqnarray}\label{hat_U}
{\hat U}_{\ell,\ell\pm1}(t) &=&
  \exp\left(-i\frac{ed}{\hbar c}A^z_{\ell,\ell\pm1}(t){\hat\tau}_3\right),\\
A^z_{\ell,\ell\pm1}(t) &=&
  \frac{1}{d}\int_{\ell\cdot d}^{(\ell\pm1)\cdot d} dz A^z(z,t),
\end{eqnarray}
where $d$ is the layer spacing.
The effective interlayer scattering lifetime, $\tau_\bot(s,s')$, is
generally anisotropic.
We describe this anisotropy by
\begin{eqnarray}
\frac{1}{\tau_\bot(s,s')}=\frac{1}{\tau_\bot}
 \frac{\exp\left[\gamma\,\cos{\left(\varphi(s)-\varphi(s')\right)}\right]}
      {\langle\langle\exp\left[
		\gamma\,\cos{\left(\varphi(s)-\varphi(s')\right)}
       \right]\rangle_s\rangle_{s'}}.
\label{tau_perp}
\end{eqnarray}
The real space angles $\varphi(s)$ and $\varphi(s')$ give the in-plane
directions of the quasiparticle Fermi velocity before and after scattering
to an adjacent layer.  The parameter $\gamma$ specifies to what degree the
scattered electrons ``remember'' their initial momentum.  Isotropic
scattering corresponds to $\gamma=0$, while extreme forward scattering
corresponds to $\gamma\to\infty$.  The scattering rate is illustrated
in figure~(\ref{polar_plot}{}) for several values of the parameter
$\gamma$.  Note that in the interlayer diffusion model one needs an
anisotropic $c$-axis scattering rate to get a nonzero interlayer Josephson
coupling with a $d$-wave order parameter.\cite{graf94}
However, {higher order tunneling processes also give a nonzero
Josephson coupling.\cite{tanaka94}}

Here we are concerned with the current response to a weak electric
field for a layered $d$-wave superconductor. The in-plane current is
carried by quasiparticles moving with an in-plane velocity ${\bf v}_f$, and
is given by standard equations of quasiclassical theory
\cite{serene83,rammer86,larkin86}
\begin{eqnarray}\label{j_xy}
{\bf j}_{\ell}(t)= eN_f
  \int{d\epsilon\over 4\pi i} \left<
         {\bf v}_f(s)\, {\mbox{Tr}} \left[{\hat\tau}_3\,
                              \hat g_\ell^K(s;\epsilon,t)\right]\right>_s
\ .
\end{eqnarray}
The interlayer current, on the other hand, is of a different nature.
It originates from random scattering processes of quasiparticles between
adjacent layers. The microscopic expression for the interlayer current
density was derived for isotropic interlayer scattering in
Ref.~\onlinecite{graf93}, and is generalized below to anisotropic scattering,
\begin{eqnarray}\label{j_z}
j^z_{\ell,\ell+1}(t) &=& -\frac{eN_f d}{i\hbar}
    \int{d\epsilon\over 4\pi i}
        \left\langle
         \mbox{Tr}\left\{ {\hat\tau}_3 \;
\Bigl[
   \mbox{$\hat{\mbox{\footnotesize
$\Sigma$}}$}_{\ell,\ell+1}^\bot(s;\epsilon,t)
\ ,
   \hat g_\ell(s;\epsilon)
\Bigr]_\otimes\right\}^K
\right\rangle_s \nonumber \\
&=&
   -\frac{eN_f d}{i\hbar} \int{d\epsilon\over 4\pi i}
        \left\langle
        {\mbox{Tr}} \left\{ {\hat\tau}_3 \;
        \Bigl( \Bigr. \right.\right.
           {\mbox{$\hat{\mbox{\footnotesize $\Sigma$}}$}_{\ell,\ell+1}^{\bot,\,
R}}\otimes {{\hat g}}_\ell^K+
           {\mbox{$\hat{\mbox{\footnotesize $\Sigma$}}$}_{\ell,\ell+1}^{\bot,\,
K}}\otimes {{\hat g}}_\ell^A
	   -{{\hat g}}_\ell^R\otimes {\mbox{$\hat{\mbox{\footnotesize
$\Sigma$}}$}_{\ell,\ell+1}^{\bot,\, K}}-
           {{\hat g}}_\ell^K\otimes {\mbox{$\hat{\mbox{\footnotesize
$\Sigma$}}$}_{\ell,\ell+1}^{\bot,\, A}}
        \left.\left.\Bigl.\Bigr)\right\} \right\rangle_s  \ .
\end{eqnarray}
Equations~(\ref{j_xy}) and~(\ref{j_z}) cover the current response for
both normal and superconducting states, in equilibrium and
nonequilibrium situations.

\subsection{{Equilibrium Solution}}

In the absence of external perturbations, the quasiclassical transport
equation and normalization condition for the equilibrium retarded and
advanced propagators become
\begin{equation}
\left[ \epsilon\hat{\tau}_3 -
       \hat{\Delta}^{R,A}\left(s\right) -
       \mbox{$\hat{\mbox{\footnotesize
$\Sigma$}}$}^{R,A}\left(s;\epsilon\right) ,
       \hat{g}^{R,A}\left(s;\epsilon\right) \right] = 0 ,
\label{uni_eq1}
\end{equation}
\begin{equation}
\left[ \hat{g}^{R,A}\left(s;\epsilon\right) \right]^{2} =
-\pi^{2}\hat{1} ,
\label{uni_norm1}
\end{equation}
where the $\otimes$ products reduce to simple matrix
multiplication.
  We dropped the layer index $\ell$, since all layers are equivalent in
equilibrium.
In the real gauge, the pairing field has the simple form,
$\hat{\Delta}^{R,A}(s) = i\Delta_0(s)\hat{\tau}_2$.
Choosing this gauge, we solve equations~(\ref{uni_eq1})
and~(\ref{uni_norm1}), and obtain
\begin{equation}
\hat{g}^{R,A}\left(s;\epsilon\right) =
  -\pi\frac{\tilde{\epsilon}^{R,A}\left(s;\epsilon\right)\hat{\tau}_{3} -
              i\tilde{\Delta}^{R,A}\left(s;\epsilon\right)\hat{\tau}_{2}}
           {\sqrt{\tilde{\Delta}^{R,A}\left(s;\epsilon\right)^{2} -
                  \tilde{\epsilon}^{R,A}\left(s;\epsilon\right)^{2}}} ,
\label{g_equil}
\end{equation}
where the $\tilde{\epsilon}$ and $\tilde{\Delta}$ can be interpreted as
the ``renormalized'' excitation energy and gap function due to
scattering processes. The renormalized quantities are determined by
the scattering self-energy, $\mbox{$\hat{\mbox{\footnotesize
$\Sigma$}}$}^{R,A}$, and are given by
\begin{eqnarray}
\tilde{\epsilon}^{R,A}\left(s;\epsilon\right) &=&
  \epsilon - {1\over 2}{\mbox{Tr}}\left[\hat{\tau}_3
             \mbox{$\hat{\mbox{\footnotesize
$\Sigma$}}$}^{R,A}\left(s;\epsilon\right)\right] , \label{epst_def} \\
\tilde{\Delta}^{R,A}\left(s;\epsilon\right) &=&
  \Delta_0\left(s\right) - {i\over 2}{\mbox{Tr}}\left[\hat{\tau}_2
             \mbox{$\hat{\mbox{\footnotesize
$\Sigma$}}$}^{R,A}\left(s;\epsilon\right)\right] . \label{delt_def}
\end{eqnarray}
The in-plane scattering self-energy can be evaluated by solving
equation~(\ref{t_matrix}) to obtain
\begin{equation}
\mbox{$\hat{\mbox{\footnotesize $\Sigma$}}$}^{R,A}_{\|}\left(\epsilon\right) =
\frac{\hbar}{2\tau_\|}
  \frac{1}
       {1 - \bar\sigma\left[1 +
      \left(\pi^{-1}\left<\hat{g}^{R,A}(s;\epsilon)\right>_s\right)^{2}\right]}
  \left[\sqrt{\frac{1}{\bar\sigma}-1} +
        \frac{\left<\hat{g}^{R,A}(s;\epsilon)\right>_s}{\pi}\right] ,
\label{sigma_equil}
\end{equation}
while the interplane self-energy has the simple form
\begin{equation}
\mbox{$\hat{\mbox{\footnotesize
$\Sigma$}}$}^{R,A}_{\bot}\left(s;\epsilon\right) = \frac{\hbar}{2\pi}
 \left<\frac{1}{\tau_\bot\left(s,s'\right)}
         \hat{g}^{R,A}(s';\epsilon)\right>_{s'} .
\label{sigma_c_equil}
\end{equation}

{}From the solutions for the equilibrium retarded and advanced propagators
and self-energies, one can calculate the equilibrium Keldysh components,
\begin{eqnarray}
\hat{g}^{K}\left(s;\epsilon\right) =
  \tanh\left(\frac{\epsilon}{2\mbox{$k_{\text{\tiny B}}$} T}\right)
  \left(\hat{g}^{R}\left(s;\epsilon\right) -
        \hat{g}^{A}\left(s;\epsilon\right)\right) , \label{g0K} \\
\mbox{$\hat{\mbox{\footnotesize $\Sigma$}}$}^{K}\left(s;\epsilon\right) =
  \tanh\left(\frac{\epsilon}{2\mbox{$k_{\text{\tiny B}}$} T}\right)
  \left(\mbox{$\hat{\mbox{\footnotesize $\Sigma$}}$}^{R}\left(s;\epsilon\right)
-
        \mbox{$\hat{\mbox{\footnotesize
$\Sigma$}}$}^{A}\left(s;\epsilon\right)\right) . \label{sigma0K}
\end{eqnarray}

\subsection{{The $d$-wave Model}}

So far the presentation of the quasiclassical theory has been fairly general.
To make further progress we must specify the shape of the Fermi surface and
the form of the pairing interaction, $V(s,s')$.
We consider a simple model that incorporates $d$-wave pairing.
We assume a cylindrical Fermi surface (in which case the Fermi surface
position,
$s$, is replaced by the azimuthal angle, $\phi$), and a pairing interaction
of the form
\begin{equation}
V\left(\phi,\phi'\right) =
  V_{0}\,\cos{\left(2\phi\right)}\cos{\left(2\phi'\right)} . \label{d-gap}
\end{equation}
Note that the basis functions for this pairing interaction possess the
usual $k_x^2-k_y^2$ anisotropy. The angular dependences of the
self-energies are now determined, and the remaining task is to solve
for the magnitudes and energy dependences. For a system in equilibrium,
and in the absence of any perturbations, we find from
equation~(\ref{pair_SE}) that the pairing field has the form
\begin{equation}
\hat{\Delta}_{0}^{R,A}\left(\phi\right) =
  i\Delta_{0}\cos{\left(2\phi\right)}\hat{\tau}_{2} \,,
\label{Delta0_d}
\end{equation}
and from equation~(\ref{sigma_equil}) the in-plane scattering
self-energy becomes
\begin{equation}
\mbox{$\hat{\mbox{\footnotesize $\Sigma$}}$}^{R,A}_{\|}\left(\epsilon\right) =
\frac{\hbar}{2\tau_\|}
  \frac{1}
    {1 - \bar\sigma\left[1 +
     \left(\pi^{-1}\langle g^{R,A}_{3}(\phi;\epsilon)\rangle_\phi
           \right)^{2}\right]}
  \left[\sqrt{\frac{1}{\bar\sigma}-1} +
        \frac{\langle g^{R,A}_{3}(\phi;\epsilon)\rangle_\phi
              \hat{\tau}_{3}}{\pi}\right] ,
\label{sigma_d}
\end{equation}
where
\begin{equation}
\langle g_{3}^{R,A}\left(\phi;\epsilon\right)\rangle_\phi =
\oint\frac{d\phi}{2\pi}
  \frac{-\pi\tilde{\epsilon}^{R,A}\left(\epsilon\right)}
                  {\sqrt{\Delta_{0}^{2}\cos^2{\left(2\phi\right)} -
                   \tilde{\epsilon}^{R,A}\left(\epsilon\right)^{2}}} .
\label{g3bar}
\end{equation}
Note that $\mbox{$\hat{\mbox{\footnotesize
$\Sigma$}}$}^{R,A}_{\|}\left(\epsilon\right)$ has no
$\hat{\tau}_{2}$ component because the Fermi surface average of
$\hat{\Delta}_{0}^{R,A}\left(\phi\right)$ is zero.
However the $\hat{\tau}_{2}$ term does not necessarily drop out of the
{\it interplane} scattering self-energy because of the anisotropy of the
interplane scattering rate, $\tau_{\perp}(s,s')^{-1}$.

In the calculations that follow we assume that the interplane scattering rate,
$1/\tau_{\perp}$, is small compared to the in-plane scattering rate,
$1/\tau_{||}$, in which case we neglect corrections of the order
$1/\tau_{\perp}$ in the renormalized self-energies and retain only the leading
contribution in $1/\tau_{\perp}$ that determines the $c$-axis current response.
The equilibrium propagator can then be written as
\begin{equation}
\hat{g}^{R,A}\left(\phi;\epsilon\right) =
-\pi\frac{\tilde{\epsilon}^{R,A}\left(\epsilon\right)\hat{\tau}_{3} -
              i\Delta_{0}\cos{\left(2\phi\right)}\hat{\tau}_{2}}
           {\sqrt{\Delta_{0}^{2}\cos^2{\left(2\phi\right)} -
                  \tilde{\epsilon}^{R,A}\left(\epsilon\right)^{2}}} ,
\label{g0_d}
\end{equation}
where the renormalized excitation energy is given in equation~(\ref{epst_def}).
Normally one eliminates the pairing interaction, $V_0$, and frequency
cutoff in favor of $T_c$, which is then a material parameter taken from
experiment. The gap parameter, $\Delta_0(T)$, is then calculated
self-consistently in terms of $T_c$ and the parameters defining the
self-energies $\mbox{$\hat{\mbox{\footnotesize
$\Sigma$}}$}^{R,A}_{\|}\left(\epsilon\right)$.  In the
absence of detailed information on the dominant inelastic processes we
have effectively made relaxation time approximations for
$\mbox{$\hat{\mbox{\footnotesize
$\Sigma$}}$}^{R,A}_{\bot,\|}\left(s;\epsilon\right)$, which are parametrized
by temperature-dependent relaxation rates, $\tau_\|^{-1}$ and
$\tau_\bot^{-1}(s,s')$. Thus, we take $\tau_\|(T)$, $\tau_\bot(T)$, and
$\Delta_0(T)$ as material parameters and calculate measurable properties
below $T_c$ in terms of these parameters and the assumed pairing
symmetry. Given a gap $\Delta_{0}$, one must solve
equations~(\ref{epst_def}) and (\ref{g0_d}) self-consistently for
$\tilde{\epsilon}^{R,A}\left(\epsilon\right)$. The equations can be
solved numerically via an iterative technique to obtain the density of states
\begin{equation}
N(\epsilon) = -\frac{N_f}{\pi}\mbox{Im}
                   \left<g_{3}^{R}\left(\phi;\epsilon\right)\right>_\phi .
\label{DOS}
\end{equation}
The equilibrium solutions for
$\tilde{\epsilon}^{R,A}\left(\epsilon\right)$ and
$\hat{g}^{R,A}\left(\phi;\epsilon\right)$ are also needed to carry out
the current response calculations described below.  Although the
density of states, for an unconventional superconductor with lines of nodes,
has been known for some time (see review~\onlinecite{sigrist91}) we find
it useful to discuss our results for the conductivity in conjunction with
the density of states.

\section{LINEAR CURRENT RESPONSE} \label{Lin_Resp}

Calculation of the infrared conductivity requires an appropriate
formulation of linear response theory. The quasiclassical formulation
of linear response has been developed by several authors ({cf.}
Ref.~\onlinecite{yip92a}). We give a brief summary of an efficient
formulation of linear response theory that utilizes the normalization
conditions and related algebraic identities to solve the linearized
quasiclassical transport equations.\cite{rainer92} We consider a
time-dependent electric field that is uniform in space. This is a good
approximation for high-$T_c$ superconductors with a penetration depth
much larger than the mean free path and the coherence length.

\subsection{{In-plane Conductivity}}

For a uniform field it is convenient to write
equation~(\ref{keldysh_trans}) in the compact form
\begin{eqnarray}\label{d_QCeq}
\displaystyle
\left [
   \epsilon {\hat\tau}_3 - \delta\hat{v} - \mbox{$\hat{\mbox{\footnotesize
$\Sigma$}}$} \, , \,
   {\hat g}
\right ]_\otimes = 0
\ ,
\end{eqnarray}
where $\delta\hat{v}$ is the perturbation, and the pairing field,
$\hat{\Delta}$, has been included in the full self-energy,
$\mbox{$\hat{\mbox{\footnotesize $\Sigma$}}$}$.
For the electrical conductivity the perturbation is of the form
\begin{eqnarray}
& & \delta\hat{v}^{R,A}(s,t) = -\frac{e}{c}{\bf v}_f(s)\cdot{\bf
A}(t)\,\hat{\tau}_3,
\label{our_perturb_RA} \\
& & \delta\hat{v}^{K}(s,t) = 0 .
\end{eqnarray}

We assume that the equilibrium solution, denoted by
$\mbox{$\hat{\mbox{\footnotesize $\Sigma$}}$}_{0}$ and
$\hat{g}_{0}$, is known, then linearize the quasiclassical transport
equations in terms of the perturbation,
$\delta\hat{v}\left(s;t\right)$, and the response,
$\delta\hat{g}\left(s;\epsilon,t\right)$.  In general these linearized
transport equations involve a self-energy response,
$\delta\mbox{$\hat{\mbox{\footnotesize $\Sigma$}}$}\left(s;\epsilon,t\right)$,
corresponding to the vertex
corrections in a Green function formalism. However, for the case of
uniform perturbations and isotropic scattering rates, the vertex
corrections to the transverse conductivity vanish.
{Note that in calculating the in-plane current response, we
neglect the contribution of the interlayer scattering self-energy to
$\mbox{$\hat{\mbox{\footnotesize $\Sigma$}}$}_{0}$.}
We write the matrix propagator in the form
\begin{equation}
\hat{g}\left(s;\epsilon,t\right) =
  \hat{g}_{0}\left(s;\epsilon\right) +
  \delta\hat{g}\left(s;\epsilon,t\right) ,
\label{delta_g}
\end{equation}
then expand equation~(\ref{d_QCeq}) to first order to obtain
the linearized transport equation,
\begin{eqnarray} \label{lin_eq1}
&{\left[\epsilon\hat{\tau}_3 -
        \mbox{$\hat{\mbox{\footnotesize
	$\Sigma$}}$}_{0}\left(s;\epsilon\right)\right]
  		\!\otimes\! \delta\hat{g}\left(s;\epsilon,t\right)  -
	\delta\hat{g}\left(s;\epsilon,t\right)\!\otimes\!
	  	\left[\epsilon\hat{\tau}_3 -
        \mbox{$\hat{\mbox{\footnotesize
	$\Sigma$}}$}_{0}\left(s;\epsilon\right)\right]
\nonumber} & \\ &
=\delta\hat{v}\left(s;t\right)\!\otimes\!
  \hat{g}_{0}\left(s;\epsilon\right) -
\hat{g}_{0}\left(s;\epsilon\right)\!\otimes\!
  \delta\hat{v}\left(s;t\right) . & \nonumber\\ &&\hspace{-20mm}
\end{eqnarray}
In addition, we expand the normalization condition through first order
and obtain
\begin{eqnarray}
\lefteqn{\hat{g}_{0}\left(s;\epsilon\right)\otimes
\hat{g}_{0}\left(s;\epsilon\right)  = -\pi^{2}\hat{1} ,}
\label{lin_norm1} \\
&  &\hat{g}_{0}\left(s;\epsilon\right)\otimes
    \delta\hat{g}\left(s;\epsilon,t\right) +
  \delta\hat{g}\left(s;\epsilon,t\right)\otimes
    \hat{g}_{0}\left(s;\epsilon\right) = 0 .
\label{lin_norm2}
\end{eqnarray}
The $\otimes$ product denotes a folding product in the energy-time
domain, together with matrix multiplication of Nambu-Keldysh matrices.
Fourier transforming the time variable turns the energy-time folding in
equations~(\ref{lin_eq1}) and~(\ref{lin_norm2}) into a simple product,
with an energy shift of magnitude $\frac{\hbar\omega}{2}$ in the equilibrium
quantities. This simplification is a consequence of the time
independence of the unperturbed propagator and self-energy. Thus,
equations~(\ref{lin_eq1}) and~(\ref{lin_norm2}) become
\begin{eqnarray}
&
\left[\left(\epsilon+\frac{\hbar\omega}{2}\right)\hat{\tau}_3 -
      \mbox{$\hat{\mbox{\footnotesize
$\Sigma$}}$}_{0}\left(s;\epsilon+\frac{\hbar\omega}{2}\right)\right]
  \otimes \delta\hat{g}\left(s;\epsilon,\omega\right)
  -\,
\delta\hat{g}\left(s;\epsilon,\omega\right)\otimes
  \left[\left(\epsilon-\frac{\hbar\omega}{2}\right)\hat{\tau}_3 -
        \mbox{$\hat{\mbox{\footnotesize
$\Sigma$}}$}_{0}\left(s;\epsilon-\frac{\hbar\omega}{2}\right)\right]
  \nonumber \\
  &  =\,
\delta\hat{v}\left(s;\omega\right)\otimes
  \hat{g}_{0}\left(s;\epsilon-\frac{\hbar\omega}{2}\right) -
\hat{g}_{0}\left(s;\epsilon+\frac{\hbar\omega}{2}\right)\otimes
  \delta\hat{v}\left(s;\omega\right) ,
\label{lin_eq1f}
\end{eqnarray}
\begin{equation}
  \hat{g}_{0}\left(s;\mbox{$\epsilon\!+\!{\hbar\omega\over 2}$}\right)\!\otimes
    \!\delta\hat{g}\left(s;\epsilon,\omega\right) +
  \delta\hat{g}\left(s;\epsilon,\omega\right)\!\otimes\!
    \hat{g}_{0}\left(s;\mbox{$\epsilon\!-\!{\hbar\omega\over 2}$}\right) = 0 ,
\label{lin_norm2f}
\end{equation}
where the $\otimes$ product in equations~(\ref{lin_norm1}),
(\ref{lin_eq1f}), and~(\ref{lin_norm2f}) should now be interpreted as a
simple matrix-product of Nambu-Keldysh matrices.

Equation~(\ref{lin_eq1f}), supplemented by equations~(\ref{lin_norm1})
and~(\ref{lin_norm2f}), amounts to 16 coupled linear equations for the 12
independent components of the Nambu-Keldysh matrix $\delta\hat g$.
These components are conveniently described by the three 2$\times$2-Nambu
matrices $\delta\hat g^R$, $\delta\hat g^A$, and $\delta\hat g^K$.
The linear equations can be solved either directly, via matrix inversion,
\cite{yip92a}
or, alternatively, by using special algebraic identities connecting the
propagators and self-energies, together with
the normalization conditions (\ref{lin_norm1}) and (\ref{lin_norm2f}).
\cite{rainer92}
The identities are based on the following relation between the
terms in the square brackets in equation~(\ref{lin_eq1f}), and the
unperturbed propagators,
\begin{equation}
\left[\epsilon\hat{\tau}_{3} -
  \mbox{$\hat{\mbox{\footnotesize
$\Sigma$}}$}_{0}^{R,A}\left(s;\epsilon\right)\right] =
C^{R,A}\left(s;\epsilon\right)
  \hat{g}_{0}^{R,A}\left(s;\epsilon\right) +
  D^{R,A}\left(s;\epsilon\right)\hat{1} ,
\label{CD_def}
\end{equation}
where $C^{R,A}\left(s;\epsilon\right)$ and
$D^{R,A}\left(s;\epsilon\right)$ are $c$-number functions that are uniquely
determined by the solutions for $\mbox{$\hat{\mbox{\footnotesize
$\Sigma$}}$}_{0}$ and $\hat{g}_{0}$.
The general solution of the linearized transport equation can be written
in terms of the unperturbed propagators, the perturbation, and the functions
$C^{R,A}$, and $D^{R,A}$,
\begin{eqnarray}
\delta\hat{g}^{R,A}\left(s;\epsilon,\omega\right) &=&
       -\frac{C_{+}^{R,A}\!\left(s;\epsilon,\omega\right)
              \hat{g}_{0}^{R,A}\!\left(s;\epsilon+\frac{\hbar\omega}{2}\right)
-
              D_{-}^{R,A}\!\left(s;\epsilon,\omega\right)}
             {\pi^{2} C_{+}^{R,A}\!\left(s;\epsilon,\omega\right)^{2} +
              D_{-}^{R,A}\!\left(s;\epsilon,\omega\right)^{2}}
\nonumber \\ & \times &
  \left[\delta\hat{v}\left(s;\omega\right)
    \hat{g}_{0}^{R,A}\left(s;\mbox{$\epsilon\!-\!{\hbar\omega\over 2}$}\right)
-
  \hat{g}_{0}^{R,A}\left(s;\mbox{$\epsilon\!+\!{\hbar\omega\over 2}$}\right)
    \delta\hat{v}\left(s;\omega\right)\right] ,
\label{dg_RA}
\end{eqnarray}
and
\begin{eqnarray}
\lefteqn{ \delta\hat{g}^{K}\left(s;\epsilon,\omega\right) =
\tanh\left(\frac{\epsilon-\frac{\hbar\omega}{2}}{2\mbox{$k_{\text{\tiny B}}$}
T}\right)
            \delta\hat{g}^{R}\left(s;\epsilon,\omega\right) -
\tanh\left(\frac{\epsilon+\frac{\hbar\omega}{2}}{2\mbox{$k_{\text{\tiny B}}$}
T}\right)
            \delta\hat{g}^{A}\left(s;\epsilon,\omega\right) }
  \nonumber \\ & & \hspace{50mm} +
\left(\tanh\left(\frac{\epsilon+\frac{\hbar\omega}{2}}{2\mbox{$k_{\text{\tiny
B}}$} T}\right) -
      \tanh\left(\frac{\epsilon-\frac{\hbar\omega}{2}}{2\mbox{$k_{\text{\tiny
B}}$} T}\right)\right)
  \delta\hat{g}^{a}\left(s;\epsilon,\omega\right) ,
\label{dg_K}
\end{eqnarray}
where we have written the Keldysh propagator, $\delta\hat{g}^K$, in terms
of $\delta\hat{g}^{R,A}$, and the ``anomalous'' response function introduced
by Eliashberg,\cite{eliashberg71} which has the solution,
\begin{eqnarray}
\delta\hat{g}^{a}\left(s;\epsilon,\omega\right) &=&
       -\frac{C_{+}^{a}\left(s;\epsilon,\omega\right)
              \hat{g}_{0}^{R}\left(s;\epsilon+\frac{\hbar\omega}{2}\right) -
              D_{-}^{a}\left(s;\epsilon,\omega\right)}
             {\pi^{2} C_{+}^{a}\left(s;\epsilon,\omega\right)^{2} +
              D_{-}^{a}\left(s;\epsilon,\omega\right)^{2}}
  \left[\delta\hat{v}\left(s;\omega\right)
    \hat{g}_{0}^{A}\left(s;\mbox{$\epsilon\!-\!{\hbar\omega\over 2}$}\right) -
  \hat{g}_{0}^{R}\left(s;\mbox{$\epsilon\!+\!{\hbar\omega\over 2}$}\right)
    \delta\hat{v}\left(s;\omega\right)\right] .
\label{dg_a}
\end{eqnarray}
The functions $C_{+}^{R,A,a}$, $D_{-}^{R,A,a}$ in equations~(\ref{dg_RA})
and~(\ref{dg_a}) are defined in terms of $C^{R,A}$ and $D^{R,A}$ of
equation~(\ref{CD_def}) as follows,
\begin{eqnarray}
C_{+}^{R,A}\left(s;\epsilon,\omega\right) &=&
  C^{R,A}\left(s;\mbox{$\epsilon\!+\!{\hbar\omega\over 2}$}\right) +
  C^{R,A}\left(s;\mbox{$\epsilon\!-\!{\hbar\omega\over 2}$}\right), \\
D_{-}^{R,A}\left(s;\epsilon,\omega\right) &=&
  D^{R,A}\left(s;\mbox{$\epsilon\!+\!{\hbar\omega\over 2}$}\right) -
  D^{R,A}\left(s;\mbox{$\epsilon\!-\!{\hbar\omega\over 2}$}\right),\\
C_{+}^{a}\left(s;\epsilon,\omega\right) &=&
  C^{R}\left(s;\mbox{$\epsilon\!+\!{\hbar\omega\over 2}$}\right) +
  C^{A}\left(s;\mbox{$\epsilon\!-\!{\hbar\omega\over 2}$}\right) , \\
D_{-}^{a}\left(s;\epsilon,\omega\right) &=&
  D^{R}\left(s;\mbox{$\epsilon\!+\!{\hbar\omega\over 2}$}\right) -
  D^{A}\left(s;\mbox{$\epsilon\!-\!{\hbar\omega\over 2}$}\right) .
\end{eqnarray}

Of central importance is the Keldysh propagator,
$\delta\hat g^K\left(s;\epsilon,\omega\right)$, since it gives directly the
linear response of observable quantities, such as the charge density and
the current density, to the perturbation, $\delta\hat{v}$.
Inserting $\delta\hat g^K$ from equation~(\ref{dg_K}), together
with $\delta\hat{v}$ from equation~(\ref{our_perturb_RA}), into
equation~(\ref{j_xy}) for the current gives the in-plane conductivity
\begin{eqnarray}
\lefteqn{ \sigma^\|\left(\omega\right) = \frac{2N_{f}e^{2}v_f^2}{i\hbar\omega}
  \int\frac{d\epsilon}{4\pi i}\int\frac{d\phi}{2\pi}\;\cos^2{\left(\phi\right)}
\Bigg\{
\tanh\left(\frac{\epsilon-\frac{\hbar\omega}{2}}{2\mbox{$k_{\text{\tiny B}}$}
T}\right)
  \frac{C_{+}^{R}\left(\phi;\epsilon,\omega\right)}
       {\pi^{2}C_{+}^{R}\left(\phi;\epsilon,\omega\right)^{2} +
               D_{-}^{R}\left(\phi;\epsilon,\omega\right)^{2}}
  \Bigg. } \nonumber \\ & & \hspace*{30mm} \times \left.
\left[
  g_{0}^{R}\left(\phi;\mbox{$\epsilon\!-\!{\hbar\omega\over 2}$}\right)
g_{0}^{R}\left(\phi;\mbox{$\epsilon\!+\!{\hbar\omega\over 2}$}\right) +
  f_{0}^{R}\left(\phi;\mbox{$\epsilon\!-\!{\hbar\omega\over 2}$}\right)
f_{0}^{R}\left(\phi;\mbox{$\epsilon\!+\!{\hbar\omega\over 2}$}\right) +
  \pi^{2}\right]
  \right. \nonumber \\ & & \hspace*{10mm} - \left.
\tanh\left(\frac{\epsilon+\frac{\hbar\omega}{2}}{2\mbox{$k_{\text{\tiny B}}$}
T}\right)
  \frac{C_{+}^{A}\left(\phi;\epsilon,\omega\right)}
       {\pi^{2}C_{+}^{A}\left(\phi;\epsilon,\omega\right)^{2} +
               D_{-}^{A}\left(\phi;\epsilon,\omega\right)^{2}}
  \right. \nonumber \\ & & \hspace*{30mm} \times \left.
\left[
  g_{0}^{A}\left(\phi;\mbox{$\epsilon\!-\!{\hbar\omega\over 2}$}\right)
g_{0}^{A}\left(\phi;\mbox{$\epsilon\!+\!{\hbar\omega\over 2}$}\right) +
  f_{0}^{A}\left(\phi;\mbox{$\epsilon\!-\!{\hbar\omega\over 2}$}\right)
f_{0}^{A}\left(\phi;\mbox{$\epsilon\!+\!{\hbar\omega\over 2}$}\right) +
  \pi^{2}\right]
  \right. \nonumber \\ & & \hspace*{10mm} + \left.
\left(\tanh\left(\frac{\epsilon+\frac{\hbar\omega}{2}}{2\mbox{$k_{\text{\tiny
B}}$} T}\right) -
      \tanh\left(\frac{\epsilon-\frac{\hbar\omega}{2}}{2\mbox{$k_{\text{\tiny
B}}$} T}\right)\right)
\frac{C_{+}^{a}\left(\phi;\epsilon,\omega\right)}
     {\pi^{2}C_{+}^{a}\left(\phi;\epsilon,\omega\right)^{2} +
             D_{-}^{a}\left(\phi;\epsilon,\omega\right)^{2}}
  \right. \nonumber \\ & & \hspace*{30mm} \times \Bigg.
\left[
  g_{0}^{A}\left(\phi;\mbox{$\epsilon\!-\!{\hbar\omega\over 2}$}\right)
g_{0}^{R}\left(\phi;\mbox{$\epsilon\!+\!{\hbar\omega\over 2}$}\right) +
  f_{0}^{A}\left(\phi;\mbox{$\epsilon\!-\!{\hbar\omega\over 2}$}\right)
f_{0}^{R}\left(\phi;\mbox{$\epsilon\!+\!{\hbar\omega\over 2}$}\right) +
  \pi^{2}\right]
  \Bigg\} . \label{condu_formula}
\end{eqnarray}
where $g_0^{R,A}$ and $f_0^{R,A}$ denote the upper-left and upper-right
elements of the equilibrium Nambu matrices $\hat{g}_0^{R,A}$, respectively.
Results for the in-plane conductivity of a $d$-wave superconductor are
discussed in section~\ref{Results}.

\subsection{{Interplane Conductivity}}

Now consider the current response for a weak electric field polarized
along the $c$-axis, $E^z=-\frac{1}{c}\partial_t A^z(t)$.
{}From equation~(\ref{j_z}), the $c$-axis current response becomes
\begin{equation}
\delta j^z(t)= -\frac{eN_f d}{i\hbar}
    \int{d\epsilon\over 4\pi i}
        \left\langle
         \mbox{Tr}\left\{ {\hat\tau}_3 \;
\Bigl[
   \delta\mbox{$\hat{\mbox{\footnotesize $\Sigma$}}$}^\bot(s;\epsilon,t)
\ ,
   \hat g_0(s;\epsilon)
\Bigr]_\otimes\right\}^K
\right\rangle_s ,
\label{d_j_z}
\end{equation}
where we neglect the response of $\hat g$ to $A^z$.
The interlayer self-energy is evaluated by expanding the gauge matrices
defined in equation~(\ref{hat_U}) to first order in the perturbation,
\begin{equation}
\delta\mbox{$\hat{\mbox{\footnotesize
$\Sigma$}}$}^\bot\left(s;\epsilon,t\right) =
  -\frac{ied}{2\pi c}
  \left<\frac{1}{\tau_\bot(s,s')}
  \left[A^z(t)\hat{\tau}_3\otimes\hat{g}_{0}(s';\epsilon) -
  \hat{g}_{0}(s';\epsilon)\otimes A^z(t)\hat{\tau}_3\right]\right>_{s'} .
\label{d_sigma_perp}
\end{equation}
Substituting equation~(\ref{d_sigma_perp}) into equation~(\ref{d_j_z}),
and Fourier transforming the time variable, we obtain the linear
response of the $c$-axis current, $\delta j^z(\omega)$.  The expression
for $\delta j^z(\omega)$ can be formally separated into a quasiparticle
term, $\delta j^z_{qp}(\omega)$, and a Cooper-pair term, $\delta
j^z_{cp}(\omega)$, or equivalently, two contributions to the
interplane conductivity,
$\sigma^\bot=\sigma^\bot_{qp}+\sigma^\bot_{cp}$,
\begin{eqnarray}\label{response_qp}
\lefteqn{
\sigma^\bot_{qp}(\omega)=\frac{N_fe^2d^2}{i\pi\hbar\omega}
   \int \frac{d\epsilon}{4\pi i}
   \int \frac{d\phi}{2\pi} \int \frac{d\phi'}{2\pi}
   \frac{1}{\tau_\bot(\phi,\phi')}
  \bigg\{
   \left[g^R_0(\phi,\mbox{$\epsilon\!+\!{\hbar\omega\over
2}$})-g^A_0(\phi,\mbox{$\epsilon\!-\!{\hbar\omega\over 2}$})\right]
}
\nonumber\\ & &\hspace*{80mm}\times
   \left[ g^K_{0}(\phi',\mbox{$\epsilon\!+\!{\hbar\omega\over
2}$})-g^K_{0}(\phi',\mbox{$\epsilon\!-\!{\hbar\omega\over 2}$}) \right]
\nonumber\\ & &\hspace*{20mm}
   {+}g^K_0(\phi,\mbox{$\epsilon\!+\!{\hbar\omega\over 2}$})\left[
g^A_{0}(\phi',\mbox{$\epsilon\!+\!{\hbar\omega\over
2}$})-g^A_{0}(\phi',\mbox{$\epsilon\!-\!{\hbar\omega\over 2}$})\right]
   {-}g^K_0(\phi,\mbox{$\epsilon\!-\!{\hbar\omega\over 2}$})\left[
g^R_{0}(\phi',\mbox{$\epsilon\!+\!{\hbar\omega\over
2}$})-g^R_{0}(\phi',\mbox{$\epsilon\!-\!{\hbar\omega\over 2}$})\right]
  \bigg\} ,
\end{eqnarray}
and
\begin{eqnarray}\label{response_cp}
\lefteqn{
\sigma^\bot_{cp}(\omega)=-\frac{N_fe^2d^2}{i\pi\hbar\omega}
   \int\frac{d\epsilon}{4\pi i}
   \int\frac{d\phi}{2\pi} \int\frac{d\phi'}{2\pi}
   \frac{1}{\tau_\bot(\phi,\phi')}
  \bigg\{ \bigg.
   \left[ f^R_0(\phi,\mbox{$\epsilon\!+\!{\hbar\omega\over
2}$})+f^A_0(\phi,\mbox{$\epsilon\!-\!{\hbar\omega\over 2}$}) \right]
}
\nonumber\\ & &\hspace*{80mm} \times
   \left[ f^K_{0}(\phi',\mbox{$\epsilon\!+\!{\hbar\omega\over
2}$})+f^K_{0}(\phi',\mbox{$\epsilon\!-\!{\hbar\omega\over 2}$}) \right]
\nonumber\\ & &\hspace*{20mm}
   {+}f^K_0(\phi,\mbox{$\epsilon\!+\!{\hbar\omega\over 2}$})\left[
f^A_{0}(\phi',\mbox{$\epsilon\!+\!{\hbar\omega\over
2}$})+f^A_{0}(\phi',\mbox{$\epsilon\!-\!{\hbar\omega\over 2}$})\right]
   {+}f^K_0(\phi,\mbox{$\epsilon\!-\!{\hbar\omega\over 2}$})\left[
f^R_{0}(\phi',\mbox{$\epsilon\!+\!{\hbar\omega\over
2}$})+f^R_{0}(\phi',\mbox{$\epsilon\!-\!{\hbar\omega\over 2}$})\right]
  \bigg\} .
\end{eqnarray}
Note that for an isotropic scattering rate
($\tau^{-1}_\bot(\phi,\phi')=\tau^{-1}_\bot$) the quasiparticle contribution
to $\sigma^\bot$ is finite, while the Cooper-pair contribution to
the conductivity vanishes in the case of a $d$-wave gap function.
This happens because the angular average of the $d$-wave gap-function,
and therefore $\langle f_0^{R,A,K}\rangle_{\phi}$,
vanishes by symmetry.
Since the zero-frequency limit of $\sigma^\bot_{cp}$ determines the
$dc$ Josephson current, {\it anisotropic} interplane scattering is necessary
to obtain a nonzero Josephson coupling.\cite{graf94}

The low-frequency limit of the imaginary part of $\sigma^\bot(T,\omega)$
determines the London penetration depth, $\lambda_\bot(T)$,
\begin{equation}
\frac{1}{\lambda_\bot^2(T)}=
\frac{4\pi}{c^2}\;\lim_{\omega\to 0} \omega\,\mbox{Im\,}{\sigma^\bot(T,\omega)}
{}.
\end{equation}
For $s$-wave pairing, we obtain a relation between the normal-state
$c$-axis conductivity, energy gap and $c$-axis penetration depth which is
equivalent to that derived for superconducting
alloys,\cite{abrikosov63}
\begin{equation}
\frac{1}{\lambda_\bot^2(T)}=
  \frac{4\pi^2}{\hbar c^2} \sigma_n^\bot(T)
  \Delta_0(T) \tanh\left(\frac{\Delta_0(T)}{2\mbox{$k_{\text{\tiny B}}$}
T}\right) .
\end{equation}
For incoherently coupled layers
${\sigma_n^\bot(T) =}{2e^2N_fd^2/\tau_\bot(T)}$.  This relation is useful
for estimating the London penetration depth from measurements of the
normal-state $dc$ resistivity and $T_c$, and it provides an important
consistency check on the interlayer tunneling model with $s$ wave
pairing.

\section{RESULTS} \label{Results}

The key complication in almost any quasiclassical calculation is the
requirement that $\hat{g}$ and $\hat{\Sigma}$ must be calculated
self-consistently.  While this is a formidable impediment to
analytic calculations, one can easily solve the self-consistent
equations using elementary numerical techniques with relatively limited
computing power.  The results presented here were obtained
from numerical calculations carried out on desktop workstations.

\subsection{Model Parameters}

Our model for the cuprate superconductors contains, as a minimal set,
five phenomenological material parameters: the transition temperature,
$T_c$, the total density of states at the Fermi level, $N_f$, the Fermi
velocity, $v_f$, and the in-plane and interplane scattering lifetimes,
$\tau_\|$ and $\tau_\bot$ for the normal state.  All of these
quantities can be deduced from normal-state measurements. In order to
accommodate $d$-wave pairing in the interlayer scattering model, we introduce
two additional material
parameters: the normalized scattering cross section, $\bar{\sigma}$,
and the interlayer scattering anisotropy parameter, $\gamma$ (see
equation~(\ref{tau_perp}). Our goal is to check the key assumptions
of the interlayer diffusion model; ({\it i})~that the superconducting
planes are Fermi liquids, and ({\it ii})~that the interlayer transport
is dominated by incoherent scattering processes.  We use normal state
data as input to calculations of superconducting properties, which then
allows us to check the consistency of our model.  We examine models
based on both $s$-wave and $d$-wave order parameters.

The best characterized high-$T_c$ material to date is Y-Ba-Cu-O. The
procedure for obtaining approximate values for the material parameters
(excluding $\bar{\sigma}$ and $\gamma$) was discussed in detail in
Ref.~\onlinecite{graf93}. Using the superconducting transition
temperature, $T_c=92K$, the Sommerfeld constant, $\gamma_s=200\mu
J/(K^2cm^3)$, the in-plane resistivity,
$\varrho_\|(T_c)=50\mu\Omega\,cm$, the normal-state Drude plasma
frequency, $\hbar\omega_p=1.5eV$, and the resistivity ratio,
$\left(\varrho_\bot/\varrho_\|\right)_{T_c}=200$, we obtain the
material parameters given in table~\ref{Model_Params}. We can then
calculate superconducting properties within our model, then check for
consistency with published experimental values.  Here we extend the
treatment of Ref.~\onlinecite{graf93} to include $d$-wave pairing.

Our results, for equilibrium supercurrents, are given in
table~\ref{Y-Ba-Cu-O_checks}. The lower bounds are estimates obtained
assuming a vanishing in-plane scattering rate, while the upper bounds
were obtained using the lifetimes given in table~\ref{Model_Params},
and $\bar\sigma=1$ for the $d$-wave model.   The $c$-axis quantities
were calculated using a moderate anisotropy parameter of $\gamma=0.5$.
The ``in-plane'' quantities ($\lambda_\|$ and $dH_{c2}^\bot/dT$) agree
well with experimental values for either an $s$-wave or a $d$-wave gap,
while quantities involving $c$-axis transport ($\lambda_\bot$ and
$dH_{c2}^\|/dT$) yield better agreement for $s$-wave pairing.   We
interpret the in-plane agreement as support of the hypothesis that the
superconducting layers of the cuprate compounds are well described by a
Fermi liquid theory (regardless of the symmetry of the order
parameter).   The poor agreement obtained for $d$-wave pairing, when
$c$-axis properties were considered, can be interpreted as a failure of
either the model of incoherently coupled layers, or the model of pure
$d$-wave pairing, or both.

In addition to Y-Ba-Cu-O, we have compared the interlayer diffusion model
with recent data on Ar-annealed Bi-Sr-Ca-Cu-O and O$_2$-annealed
Pb-Bi-Sr-Ca-Cu-O;\cite{kleiner94,mueller94,mueller94b} penetration depth,
$\lambda_\bot$, critical voltage, $V_c$, and critical current, $I_c$,
measured at $4.2K$.
The data is insufficient to carry out a complete analysis, but we can check
the consistency of the theory with regard to the $c$-axis properties.
Two independent relations can be derived for the $c$-axis transport
properties at low temperatures, $T\ll T_c$,
\begin{equation}
\frac{8\pi e}{\hbar c^2} \lambda_\bot^2 j_c^\bot d = 1 , \label{R_I}
\end{equation}
\begin{equation}
\frac{4\pi^2}{\hbar c^2} \frac{\lambda_\bot^2\Delta_0}{\varrho_n^\bot} =
\left\{ \begin{array}{cc}
             1,   & \;\;\;\;\mbox{for $s$-wave pairing,} \\
             R_d, & \;\;\;\;\mbox{for $d$-wave pairing,} \\
          \end{array} \right.  \label{R_II}
\end{equation}
where $d$ is the layer spacing, and $R_{d}$ is a dimensionless number.

Equation~(\ref{R_I}) relates the interlayer Josephson critical current
to the penetration depth and is independent of the pairing symmetry.
The second relation~(\ref{R_II}) is sensitive to the symmetry of the order
parameter and depends critically on the interplane anisotropy parameter,
$\gamma$, in the case of $d$-wave pairing.
In the limit of isotropic scattering ($\gamma\to0$) $R_d\to\infty$; however,
$R_d$ is finite for anisotropic scattering ($\gamma\neq0$), and can be
$\sim1$ for extreme forward scattering ($\gamma\to\infty$).
Combining equations~(\ref{R_I}) and~(\ref{R_II}) yields the
Ambegaokar-Baratoff relation\cite{ambegaokar63}
\begin{equation}
\frac{2e}{\pi\Delta_0} \varrho_n^\bot j_c^\bot d =
\left\{ \begin{array}{cc}
             1,        & \;\;\;\;\mbox{for $s$-wave pairing,} \\
             R_d^{-1}, & \;\;\;\;\mbox{for $d$-wave pairing,} \\
          \end{array} \right.  \label{R_III}
\end{equation}
which is independent of $\gamma$ for $s$-wave pairing, but not for the
$d$-wave model.   Taking values for $\lambda_\bot$, $j_c^\bot$, and
$\varrho_n^\bot\approx V_c^\bot/j_c^\bot d$, from the
experimental data, and estimating $\Delta_0$ from $T_c$ and the
clean-limit weak-coupling relations; $\Delta^{(s)}_0=1.76\mbox{$k_{\text{\tiny
B}}$} T_c$,
$\Delta^{(d)}_0=2.14\mbox{$k_{\text{\tiny B}}$} T_c$, we can evaluate the
ratios given in
equations~(\ref{R_I})-(\ref{R_III}) for the Bi-Sr-Ca-Cu-O compounds.  The
results are summarized in table~\ref{Bi-Sr-Ca-Cu-O_checks}.   The data on
Ar-annealed Bi-Sr-Ca-Cu-O are consistent with $s$-wave pairing, and also with
$d$-wave pairing provided the interplane tunneling is extremely
anisotropic ($\gamma\agt5$).   Neither the $s$-wave nor $d$-wave model
accounts for the O$_2$-annealed data; thus, the interlayer transport
mechanism in these compounds is not well described by incoherent
tunneling.   Such systems may be better described by resonant tunneling
via impurity states localized between the CuO$_2$
layers,\cite{halbritter93} or perhaps other transfer mechanisms.
\cite{bulaevskii73,kats69,frick90,liu92,kumar90,rojo94,chakravarty93}

The analysis of the equilibrium properties implies that Fermi liquid
theory provides a sound basis for the description of the in-plane
properties of high-$T_c$ superconductors, independent of the symmetry
of the superconducting order parameter.  The interplane properties, on
the other hand, depend strongly on the gap symmetry.   The experimental
situation is also not simple; results differ depending on the compound
and material preparation. For some samples we find reasonable agreement
with the universal relations implied by the interlayer diffusion
model, while other samples display significant deviations.   These
discrepancies are present for both $s$-wave and $d$-wave models, and
suggest that the model of incoherently coupled layers is an
oversimplification of the $c$-axis charge transfer mechanism in certain
high-$T_c$ materials.  Alternatives are, for instance, resonant
coupling, coupling by pair tunneling, coherent coupling, or a
combination of different types of coupling mechanisms.

\subsection{Conductivity}

The procedure used to solve equations~(\ref{epst_def}) and~(\ref{g0_d})
for $\tilde{\epsilon}^{R,A}$, which then yields the equilibrium
propagator and self-energy, was described at the end of section~II.  In
solving for the equilibrium solutions we neglect the interlayer
self-energy term. This is justified because the interlayer scattering
rate is much smaller than the in-plane scattering rate. To calculate
the in-plane conductivity, the remaining step is to carry out the
integrations in equation~(\ref{condu_formula}). A key feature of the
$d$-wave model is the presence of low-energy excitations near the nodes
of the gap. The phase space for the $d$-wave model is similar to that of
the three-dimensional polar ($p$-wave) model with a line of nodes. We
note that there is a good qualitative agreement between the infrared
conductivity calculated for a three-dimensional polar state,
\cite{klemm88,hirschfeld89} and the two-dimensional $d$-wave state
considered here. Both states possess lines of nodes in the gap.

In figure~(\ref{clean_sigma}{a}) we show the density of states for a
$d$-wave order parameter in a fairly clean sample (i.e.,
$\alpha=\hbar/2\pi\tau\Delta_0=0.01$), for several values of the
normalized cross section.  The key feature to note is the enhancement
of the density of states at low frequencies.
\cite{schmitt-rink86,hirschfeld86,sigrist91}
This feature is strongest in the unitarity limit, and softens as one goes to
lower values of $\bar{\sigma}$, until it vanishes completely in the
Born limit.  This enhancement is interpreted as being due to
bound-states, localized at the scattering centers, whose origin lies in
local pair-breaking effects.\cite{preosti94,choi94} At a finite
concentration of scattering centers, the bound states overlap and form
``bound bands''.

The bound states are optically active; the matrix elements coupling the
impurity bound states do not vanish as $q\rightarrow0$.   This is
evident from figure~(\ref{clean_sigma}{b}), where we show the in-plane
conductivity for a sample with the same parameters as in
figure~\ref{clean_sigma}{a}), at $T=0$. In the unitarity limit
($\bar\sigma\rightarrow1$) the absorption is substantially enhanced at
low frequencies compared to the Born limit
($\bar\sigma\rightarrow0$).   The enhancement is due to transitions
within the bound band.  Note that the enhancement of the conductivity
at low frequencies, which shows the strong optical activity of the
bound states, is much more pronounced than the enhancement in the
density of states.  Furthermore, the in-plane conductivity approaches
the universal limit obtained by Lee,\cite{lee93}
$\lim_{\omega\rightarrow
0}\sigma^{||}(\omega)\simeq N_fe^2v_f^2 \hbar/\pi\Delta_0$, independent of the
scattering cross section ($\bar{\sigma}$), as shown in
figure~(\ref{clean_sigma}{b}), and scattering rate ($\alpha$).  In
addition to the strong absorption at low frequencies there is also a
small feature at $\hbar\omega\approx\Delta_0$, which corresponds to the
excitation of a quasiparticle from a bound state to the continuum
states at energies $\hbar\omega\sim\Delta_0$. Finally, we note that
there is no discernible feature at $\hbar\omega=2\Delta_0$, in contrast
to the case of $s$-wave pairing.  Thus, optical experiments are not
expected to provide a good means of measuring the maximum gap in
$d$-wave superconductors.

In figures~(\ref{dirty_sigma}{a}) and~(\ref{dirty_sigma}{b}) we show
the density of states and in-plane conductivity for a sample with a
larger scattering rate (i.e., $\alpha=\hbar/2\pi\tau\Delta_0=0.1$).  In
this case the bound states have been merged with the continuum states
and are no longer distinctly visible in the density of states.  However
the effects of the bound states can still be seen in the infrared
conductivity. Note the different scale between
fig.~(\ref{clean_sigma}{b}) and fig.~(\ref{dirty_sigma}{b}) is due to
the change in the normal-state Drude conductivity. The universal value
of the conductivity for the superconducting state implies
$\lim_{\omega\rightarrow
0}\sigma^{||}(\omega)/\sigma^{||}_{n}(0)\simeq 2\alpha$. Note also that the
absorption at low frequencies is still much larger for resonant
scattering ($\bar\sigma\to1$) than it is in the Born limit
($\bar\sigma=0$).

In figure~(\ref{Tdep_sigma}{a}) we show the temperature dependence of
the in-plane conductivity for the unitarity limit and
$\alpha=\hbar/2\pi\tau\Delta_0=0.01$ as in
figure~(\ref{clean_sigma}{}). Typically the scattering rate,
$\tau^{-1}$, will itself depend on temperature.  However, this
temperature dependence, which is obtained from experimental
measurements in our model, has been neglected here.  Note that the
gap-like feature at $\hbar\omega\approx\Delta_0$ is strongest at low
temperatures, and can be as much as 50\% higher than the normal-state
Drude value of the conductivity.  At higher temperatures we find an
upturn in the conductivity at low frequencies, which yields a coherence
peak (i.e., the conductivity exceeds the
normal state value).

The $c$-axis transport calculations were
carried out with a moderate value of the scattering anisotropy
parameter, $\gamma=0.5$ (see equation~(\ref{tau_perp}), unless
otherwise stated. In figures~(\ref{clean_sigma}{c})
and~(\ref{dirty_sigma}{c}) we show results for the real part of the
$c$-axis conductivity at $T=0$, for two values of the scattering
lifetime, and for both $s$-wave and $d$-wave pairing. Note the onset
of dissipation for isotropic $s$-wave pairing at $\hbar\omega=2\Delta_0$.
It is interesting to note that although the dynamical tunneling
currents, for the case of an $s$-wave gap, obey the frequency dependence
of ``dirty'' superconductors, their dependence on the scattering
lifetime is just the inverse (i.e., $\sigma^\bot\propto\tau_\bot^{-1}$).
Thus, in the interlayer diffusion model, an increase in interlayer
scattering increases the $c$-axis charge transfer (regardless of the
order parameter symmetry), whereas it reduces the charge transfer in
``dirty'' alloys. It is also significant that the $c$-axis conductivity in the
interlayer diffusion model does not tend to a universal value in the limit
$\omega\rightarrow 0$.

A $d$-wave superconductor has residual dissipation even at $\omega=0$,
due to low energy excitations near the nodes in the gap, which is
significantly enhanced in the limit of strong scattering.  The knee
that develops in the limit of resonant in-plane scattering
($\bar\sigma\to1$) at $\hbar\omega\approx\Delta_0$ is due to empty bound
states, as discussed for the in-plane conductivity.  Thus, it may be
possible to shed light on the strength of in-plane scattering by
detecting this common feature in both conductivities.
In figure~(\ref{Tdep_sigma}{b}) we show the temperature behavior of
the $c$-axis conductivity for the unitarity limit, and
$\alpha=\hbar/2\pi\tau\Delta_0=0.01$.  It saturates at zero-frequency
and has a washed-out minimum, which shifts slightly with increasing
temperature to higher frequencies.

\subsection{Reactance and Low-frequency Plasma mode}

The imaginary part of the conductivity (reactance) for an $s$-wave gap
is well described by a term of the form
$\sigma^{\perp}_{2}(\omega)\simeq A/\pi\omega$ for
$0\leq\hbar\omega\alt 2\Delta_0$, where $A$ is determined by the {dc}
penetration depth, $\sqrt{A}=c/2\lambda_{\perp}$. This
quasihydrodynamic approximation for the conductivity is valid at
sufficiently low frequency, when quasiparticle pair production is
negligible and the current response is dominated by the London
supercurrent. For an $s$-wave gap at $T=0$ pair production is absent for
$\hbar\omega\alt 2\Delta_0$, so the London limit is a good
approximation even for $\hbar\omega\simeq\Delta_0$.  However, the
reactance for a $d$-wave gap deviates from the $\omega^{-1}$ behavior
at frequencies much smaller than the gap, dropping much faster than the
$s$-wave result, and crossing zero at $\hbar\omega\sim 0.3\Delta_0$ (see
figure~(\ref{reactance}{}).  The breakdown of the London limit
develops at much lower frequencies for the $d$-wave gap because
quasiparticle pair production is present at any frequency for a gap
with nodes.
The fraction of the Fermi surface for which pair production
is energetically allowed is
$\phi_{\omega}\simeq\left(\hbar\omega/\pi\Delta_0\right)$. Pair
production leads to a reduction in the supercurrent, i.e., a
softening of the condensate, and of course to dissipation at all
frequencies below the maximum gap of $2\Delta_0$.
The softening of the
condensate current response appears as a reduction in the Cooper pair
contribution, $\sigma^{\perp}_{cp}$, to the conductivity compared to
the quasiparticle part, $\sigma^\bot_{qp}$.  The dashed and dotted
lines in figure~(\ref{reactance}{}) correspond to the condensate and
quasiparticle contributions to the reactance, respectively.  For the
case of an $s$-wave order parameter the condensate contribution has the
highest weight at least through frequencies on the order of the gap
amplitude, while the reactance for a $d$-wave superconductor is
dominated by quasiparticle excitations at frequencies much smaller than
the gap.

This softening of the condensate response at frequencies below
$\sim\Delta_0$, by the quasiparticle excitations, has important
implications for the low-frequency plasma mode in strongly anisotropic
superconductors.  The large anisotropy of the penetration depth in the
cuprate superconductors ($\lambda^2_\bot/\lambda^2_\|\gg 1$) leads to a
low-lying plasma mode along the $c$-axis with frequency
$\omega_{pc}\alt\omega_p\,\lambda_\|/\lambda_\bot$,\cite{woelfle94}
where $\omega_p$ is the in-plane Drude plasma frequency.
  The anisotropy of the cuprates is sufficiently large so that the collective
mode lies below $2\Delta_0$.\cite{woelfle94}  The transverse dielectric
constant, $\varepsilon_\bot(\omega)=\varepsilon_{\infty}+4\pi
i\sigma^\bot(\omega)/\omega$, vanishes at $\omega=\omega_{pc}$; thus,
for low damping the plasma frequency is obtained from the root of the
real part of $\varepsilon_\bot(\omega)$.  The frequency, $\omega^*$, at
which $\mbox{Im\,}{\sigma^\bot(\omega)}$ crosses zero provides an upper bound
for the plasma frequency, $\omega_{pc}\le\omega^*$.  Shown in the
insert of fig.~(\ref{reactance}{}) is a plot of
$\mbox{Im\,}{\sigma^\bot(\omega)}$ for both the $s$-wave and $d$-wave models
with the same zero-temperature penetration depth, $\lambda_\bot(0)$
(that is, we scaled $\mbox{Im\,}{\sigma^\bot_{(d)}(\omega)}$ of the $d$-wave
model so that
$\mbox{Im\,}{\sigma^\bot_{(d)}(0)}=\mbox{Im\,}{\sigma^\bot_{(s)}(0)}$).
Since the root, $\omega^*$, of the reactance is maximal for the clean
limit we only display the result for $\alpha\to 0$.  In the $d$-wave
case we can deduce an upper bound for $\omega_{pc}$ from the intercept
of the reactance with the $\omega$-axis, which for $\gamma=0.5$ falls
below the range of experimentally observed values
\cite{tamasaku92,homes93,muenzel94,kim94a} (denoted by the shaded
region).
Note, that $\omega_{pc}$ increases with increasing $\gamma$,
but is always below the $s$-wave result. We conclude from our analysis
of the reactance that the London approximation for the $c$-axis
conductivity (and dielectric function $\varepsilon(\omega)$) breaks
down in the case of an incoherently coupled stack of two-dimensional
$d$-wave superconductors, if the interlayer scattering is moderately
anisotropic ($\gamma\alt 0.5$).  Finally, we note that the
pair production and softening of the condensate at relatively low
frequencies leads to an anomalous temperature behavior of the surface
resistance, which increases with decreasing temperatures well below
$\Delta_0$.\cite{graf94c}

\section{CONCLUSION} \label{Conclu}

We have calculated the in-plane and $c$-axis infrared conductivities
for a layered $d$-wave superconductor within the Fermi liquid theory of
superconductivity, including the effects of in-plane and interplane
scattering.  An important effect on in-plane scattering, for $d$-wave
superconductors, is the formation of quasiparticle bound states that
are localized at the scattering centers.  These bound states are clearly
visible in the density of states, but they have a far greater effect on
the infrared conductivity (both in-plane and $c$-axis) because the
localization of these states at impurities leads to a large optical
activity.  The effects of the bound states are most pronounced for
resonant scattering centers; Zn impurities in Y-Ba-Cu-O are argued to be
candidates for resonant scatterers.\cite{pines94,zhang93}
Observation of the conductivity from impurity states in the superconducting
state of high-$T_c$ materials could provide important evidence for $d$-wave
pairing.

Our calculations of the interplane conductivity also show that the
transport properties of a layered $d$-wave superconductor are very
sensitive to the nature of the interlayer coupling.  For instance,
incoherent interlayer coupling, mediated through isotropic scattering
processes, yields a vanishing Josephson coupling because the angular
average of the $d$-wave gap function vanishes for symmetry reasons;
however, for highly anisotropic scattering (conserved in-plane momentum)
the Josephson coupling becomes of the order of the $s$-wave value.
As a result, it is difficult to reconcile the model of incoherently coupled
layers, together with $d$-wave pairing, with the currently available
$c$-axis transport data.
However, $s$-wave pairing in the interlayer diffusion model often yields
a reasonable fit to the $c$-axis data.  Thus, a better understanding of
the microscopic nature of the interplane transport mechanism is needed
in order to use $c$-axis transport measurements as probes for the order
parameter symmetry in high-$T_c$ superconductors.

\acknowledgments

 Authors M.J.G. and D.R. were supported, in part, by the ``Graduiertenkolleg
--- Materialien und Ph\"anomene bei sehr tiefen Temperaturen'' of the DFG.
  The research of M.P. was supported by the Alexander von Humboldt-Stiftung.
  J.A.S acknowledges partial support by the Science and Technology Center
for Superconductivity through NSF Grant No. 91-20000.

\appendix
\section*{}

  The study of nonequilibrium properties of a Fermi liquid, requires
knowledge of the dynamics of both the distribution of low-energy excitations
as well as the spectrum of these states.
  This may be accomplished within a framework introduced by
Keldysh.\cite{keldysh64}
  This formulation allows us to treat the spectrum and distribution function
on the same footing, and to obtain the dynamics from the standard array of
Green's function methods.
  We provide a brief summary of the definitions for the Green's functions
used in this paper, as well as the microscopic definitions of the
quasiclassical propagators.

  The two-component Nambu-field operator in layer $\ell$ is defined as
$\Psi_\ell^{\dagger}\left({\bf x},t\right) =
  \left( {\psi^{\dagger}_{\ell}}_{\uparrow}\left({\bf x},t\right) ,
         {\psi_{\ell}}_{\downarrow}\left({\bf x},t\right) \right)$
for spin singlet superconductors.
  The retarded, advanced, and Keldysh Green's functions are defined as
follows:\cite{serene83}
\begin{eqnarray}
\hat{G}^R_\ell\left({\bf x},t;{\bf x}\,',t'\right) &=&
  -\frac{i}{\hbar}\theta(t-t')\left<\left\{
    \Psi_\ell\left({\bf x},t\right) ,
    \Psi^{\dagger}_\ell\left({\bf x}\,',t'\right) \right\}\right> , \nonumber\\
\hat{G}^A_\ell\left({\bf x},t;{\bf x}\,',t'\right) &=&
  \frac{i}{\hbar}\theta(t'-t)\left<\left\{
    \Psi_\ell\left({\bf x},t\right) ,
    \Psi^{\dagger}_\ell\left({\bf x}\,',t'\right) \right\}\right> , \nonumber\\
\hat{G}^K_\ell\left({\bf x},t;{\bf x}\,',t'\right) &=&
  -\frac{i}{\hbar}\left<\left[
    \Psi_\ell\left({\bf x},t\right) ,
    \Psi^{\dagger}_\ell\left({\bf x}\,',t'\right) \right]\right> .
\end{eqnarray}
  The ``hat'' denotes a $2\times2$ matrix in particle-hole (Nambu) space.
  These propagators contain information on the physical quantities
of interest.
  The diagonal component of $\hat{G}^{R,A,K}_\ell$ describes the spectrum and
distribution of quasiparticle excitations, while its off-diagonal
component, $\hat{F}^{R,A,K}_\ell$, describes the pairing correlations.

  The quasiclassical propagators, $\hat{g}^{R,A,K}_\ell$, are obtained from the
microscopic propagators, $\hat{G}^{R,A,K}_\ell$, by integrating-out the short
wavelength, high-energy information.
  In the mixed (${\bf p},{\bf R}$) representation, the formal definition of
$\hat{g}^{R,A,K}_\ell$ is
\begin{equation} \label{xi_int}
\hat{g}^{R,A,K}_\ell\left(s,{\bf R};\epsilon,t\right) =
  \frac{1}{a}\int_{-\hbar\omega_c}^{+\hbar\omega_c}\!\! d\xi_{{\bf p}} \;\,
  \hat{\tau}_3 \hat{G}^{R,A,K}_\ell\left({\bf p},{\bf R};\epsilon,t\right) ,
\end{equation}
where $a$ is the quasiparticle renormalization factor.\cite{serene83}
  In this paper we are interested in spatially homogeneous systems in which
the $\ell$ and ${\bf R}$ dependence drops out of equation~(\ref{xi_int}).
  Thus, our propagators are functions of only the Fermi surface position
$s$, the quasiparticle excitation energy, $\epsilon$, and time $t$.
  In the transformation to the mixed representation, the Green's function
products appearing in the Dyson equation become energy-time folding
products which we denote by the symbol $\otimes$.
  A useful formal representation of this folding product is:
\begin{equation} \label{otimes}
\hat{a}(s;\epsilon,t)\otimes\hat{b}(s;\epsilon,t) =
  e^{\frac{i\hbar}{2}(\partial_\epsilon^a\partial_t^b -
                      \partial_t^a\partial_\epsilon^b)}
  \hat{a}(s;\epsilon,t)\hat{b}(s;\epsilon,t) ,
\end{equation}
  where the symbol $\partial_\epsilon^a$ indicates differentiation of
$\hat{a}$ with respect to $\epsilon$, etc.
  Note that $\hat{a}$ and $\hat{b}$ in equation~(\ref{otimes}) are
Nambu-Keldysh matrices as defined in equation~(\ref{g_nambu_keldysh}).

\twocolumn\newpage
%
%
%

\onecolumn \newpage
%
%

\centerline{
\begin{minipage}{0.6\hsize}
\begin{table}
\caption{Microscopic material parameters (Y-Ba-Cu-O).}
\label{Model_Params}
\begin{tabular}{lrrrr}
    \normalsize $T_c$
    & \normalsize $N_f$
    & \normalsize $v_{f}$
    & \normalsize $\hbar/\tau_\bot |_{T_c}$
    & \normalsize $\hbar/\tau_\| |_{T_c}$  \\
    \normalsize (K)
    & \normalsize 1/(eV\, cell\, spin)
    & \normalsize (km/s)
    & \normalsize (meV)
    & \normalsize (meV)    \\
    \hline \\
    \normalsize 92
    & \normalsize 4.4
    & \normalsize 110
    & \normalsize 0.6
    & \normalsize 14
\end{tabular}
\end{table}
\end{minipage}
}


\centerline{
\begin{minipage}{0.6\textwidth}
\begin{table}
\caption{Consistency checks (Y-Ba-Cu-O).}
\label{Y-Ba-Cu-O_checks}
\begin{tabular}{lrrrr}
   & \normalsize $\lambda_\| (0)$
   & \normalsize $\lambda_\bot (0)$
   & \normalsize ${dH_{c2}^{\|} / dT}|_{T_c}$
   & \normalsize ${dH_{c2}^{\bot} / dT}|_{T_c}$ \\
   & \normalsize (${\rm nm}$)
   & \normalsize (${\rm \mu m}$)
   & \normalsize (T/K)
   & \normalsize (T/K) \\
   \hline \\
   \normalsize Measured
   & \normalsize $\sim140$
   & \normalsize $\sim1$
   & \normalsize $\sim -10$
   & \normalsize $\sim -1.8$ \\
   \normalsize s-wave OP
   & \normalsize $130 - 160$
   & \normalsize 1.1
   & \normalsize $-14$
   & \normalsize $-2.2$ \\
   \normalsize d-wave OP
   & \normalsize $130 - 230$
   & \normalsize $7 - 10$
   & \normalsize $-55$
   & \normalsize $-1.9$
\end{tabular}
\end{table}
\end{minipage}
}


\centerline{
\begin{minipage}{0.6\hsize}
\begin{table}
\caption{Consistency checks (Bi-Sr-Ca-Cu-O).}
\label{Bi-Sr-Ca-Cu-O_checks}
\begin{tabular}{lrr}
   \normalsize Relation:
   & \normalsize  Bi-Sr-Ca-Cu-O
   & \normalsize  Pb-Bi-Sr-Ca-Cu-O \\
   & \small Ar-annealed
   & \small O$_2$-annealed \\
   \hline \\
   \normalsize $\displaystyle\frac{4\pi^2}{\hbar c^2} \frac{\lambda^2_\bot
\Delta_0} {\varrho_n^\bot}$
   & \normalsize $\sim1.5$
   & \normalsize $\sim5.9$ \\ \\
   \normalsize $\displaystyle\frac{8\pi e}{\hbar c^2}\lambda^2_\bot j_c^\bot d$
   & \normalsize $\sim1.0$
   & \normalsize $\sim0.20$ \\ \\
   \normalsize $\frac{2e}{\pi\Delta_0}\varrho_n^\bot j_c^\bot d$
   & \normalsize $\sim0.67$
   & \normalsize $\sim0.033$
\end{tabular}
\end{table}
\end{minipage}
}

\newpage
%
%

\begin{minipage}[t]{0.45\textwidth}
\begin{figure}
\epsfxsize=\hsize
\epsfbox{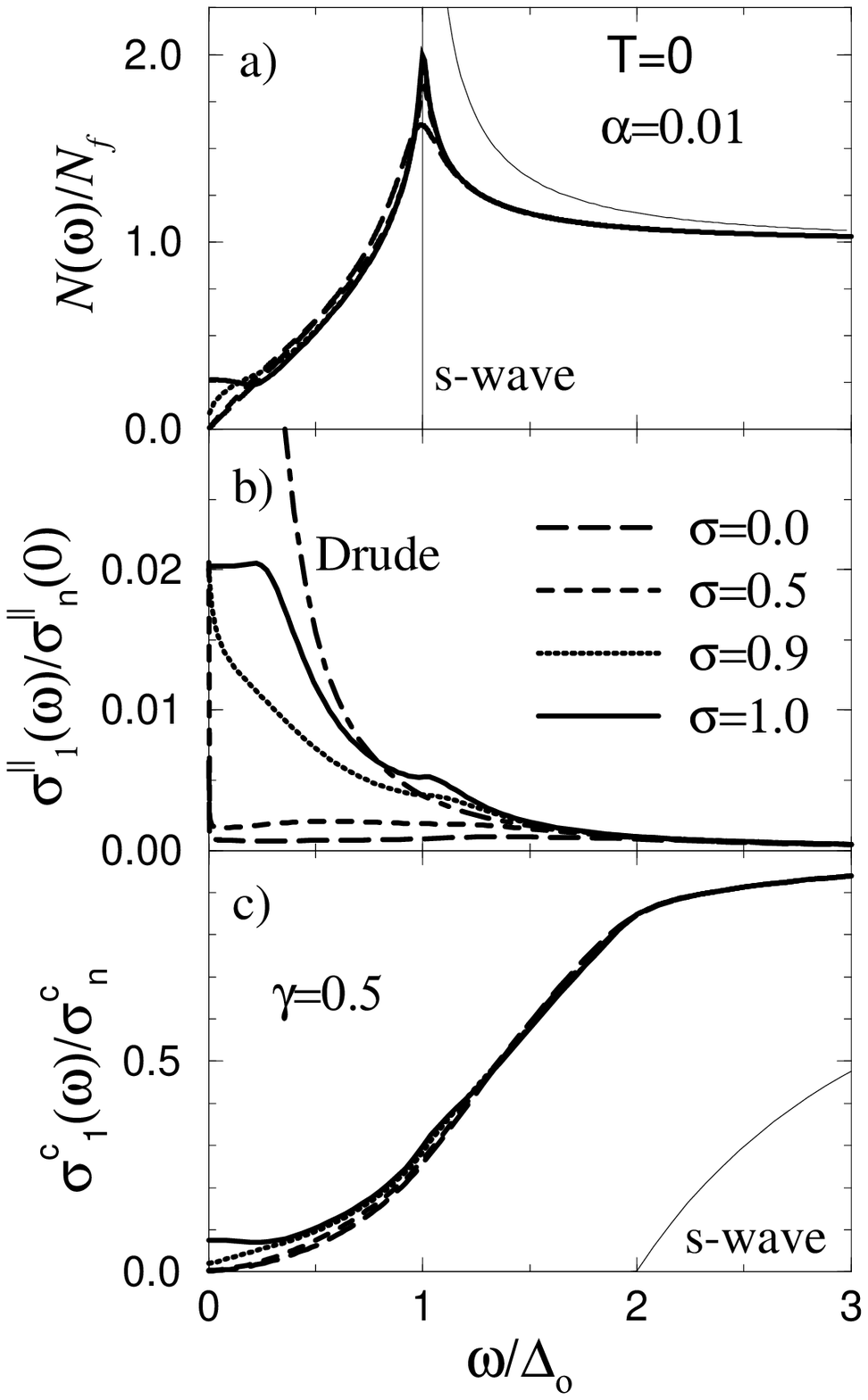}
\caption{
  The density of states and the real part of the in-plane and $c$-axis
conductivities at $T=0$, for a scattering rate described by
$\alpha=\hbar/2\pi\tau\Delta_0=0.01$, and several values of the normalized
scattering cross section $\bar\sigma$, as a function of frequency.
  All results are for a $d$-wave pairing interaction unless otherwise noted.
  (a)~Density of states normalized by the density of states at the Fermi level.
  (b)~The in-plane conductivity normalized by the normal state Drude value at
     $\omega=0$.  The corresponding $s$-wave result (not shown) becomes
     nonzero at $\hbar\omega/\Delta_0=2$ and essentially follows the Drude
     curve.
  (c)~The $c$-axis conductivity normalized by the normal state value.
     }
\label{clean_sigma}
\end{figure}
\end{minipage}
\hfill
\begin{minipage}[t]{0.45\textwidth}
\begin{figure}
\epsfxsize=\hsize
\epsfbox{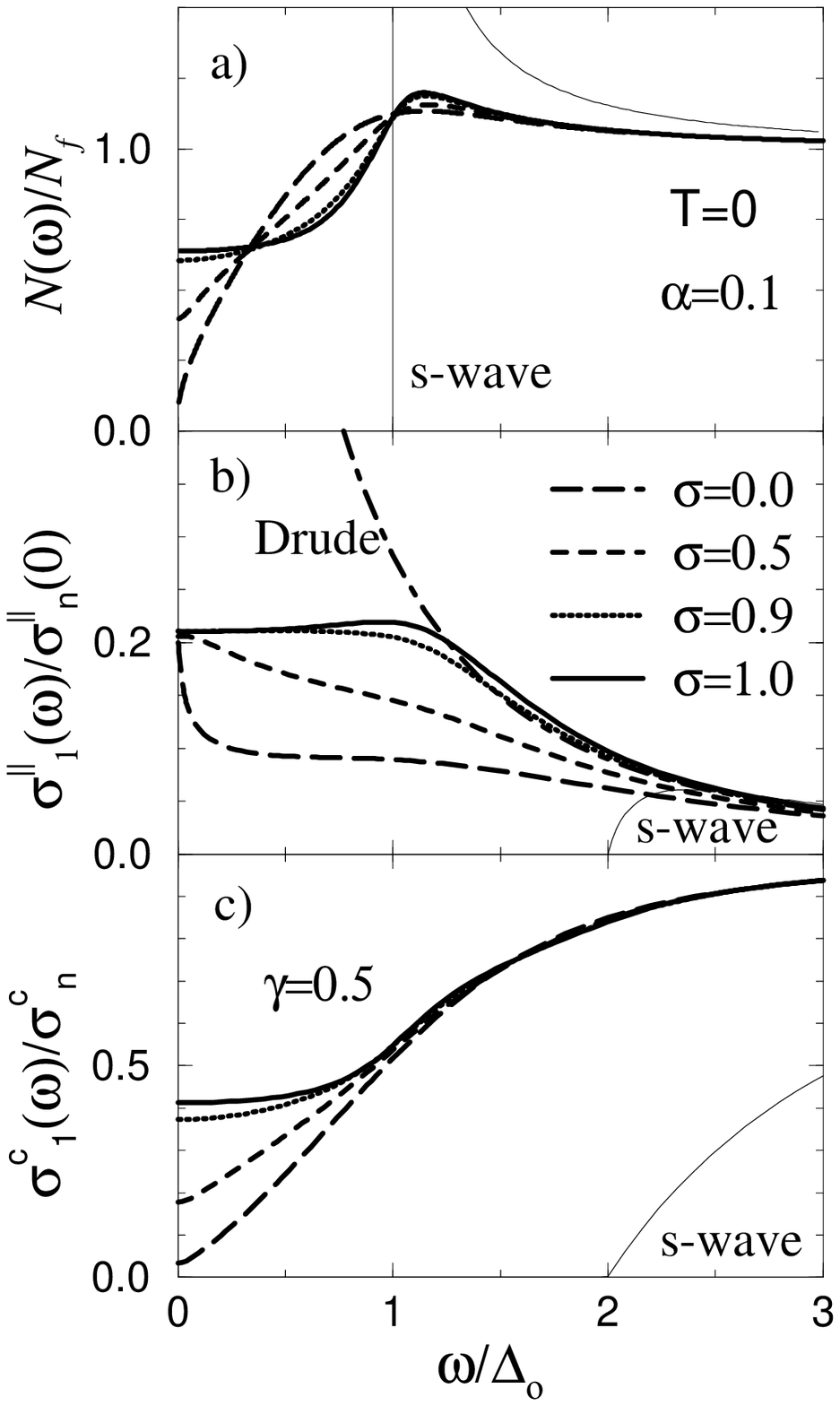}
\caption{
  The density of states and the real part of the in-plane and $c$-axis
conductivities at $T=0$, for a scattering rate described by
$\alpha=\hbar/2\pi\tau\Delta_0=0.1$, and several values of the normalized
scattering cross section $\bar\sigma$, as a function of frequency.
  All results are for a $d$-wave pairing interaction unless otherwise noted.
  (a)~Density of states normalized by the density of states at the Fermi level.
  (b)~The in-plane conductivity normalized by the normal state Drude value at
     $\omega=0$.
  (c)~The $c$-axis conductivity normalized by the normal state value.
     }
\label{dirty_sigma}
\end{figure}
\end{minipage}
\bigskip

\begin{minipage}{0.8\hsize}
\begin{figure}
\epsfxsize=\hsize
\epsfbox{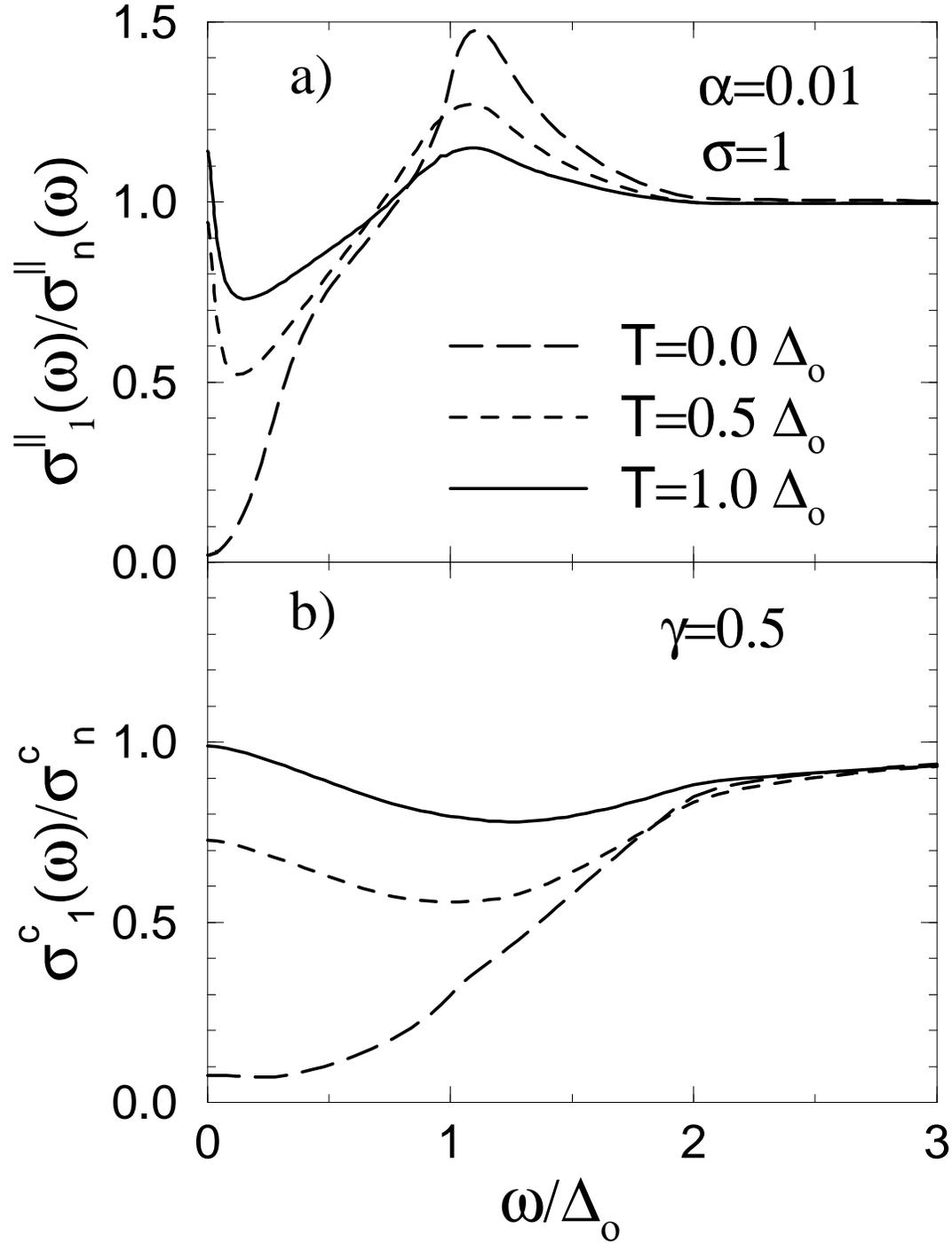}
\caption{
  The temperature dependence of the in-plane and $c$-axis conductivities for
a scattering rate of $\alpha=\hbar/2\pi\tau\Delta_0=0.01$, in the
unitarity limit, as a function of frequency.
  a)~The in-plane conductivity normalized by the normal-state Drude
     conductivity.
  b)~The $c$-axis conductivity normalized by the normal-state value.}
\label{Tdep_sigma}
\end{figure}
\end{minipage}

\begin{minipage}[b]{0.8\hsize}
\begin{figure}
\epsfxsize=\hsize
\epsfbox{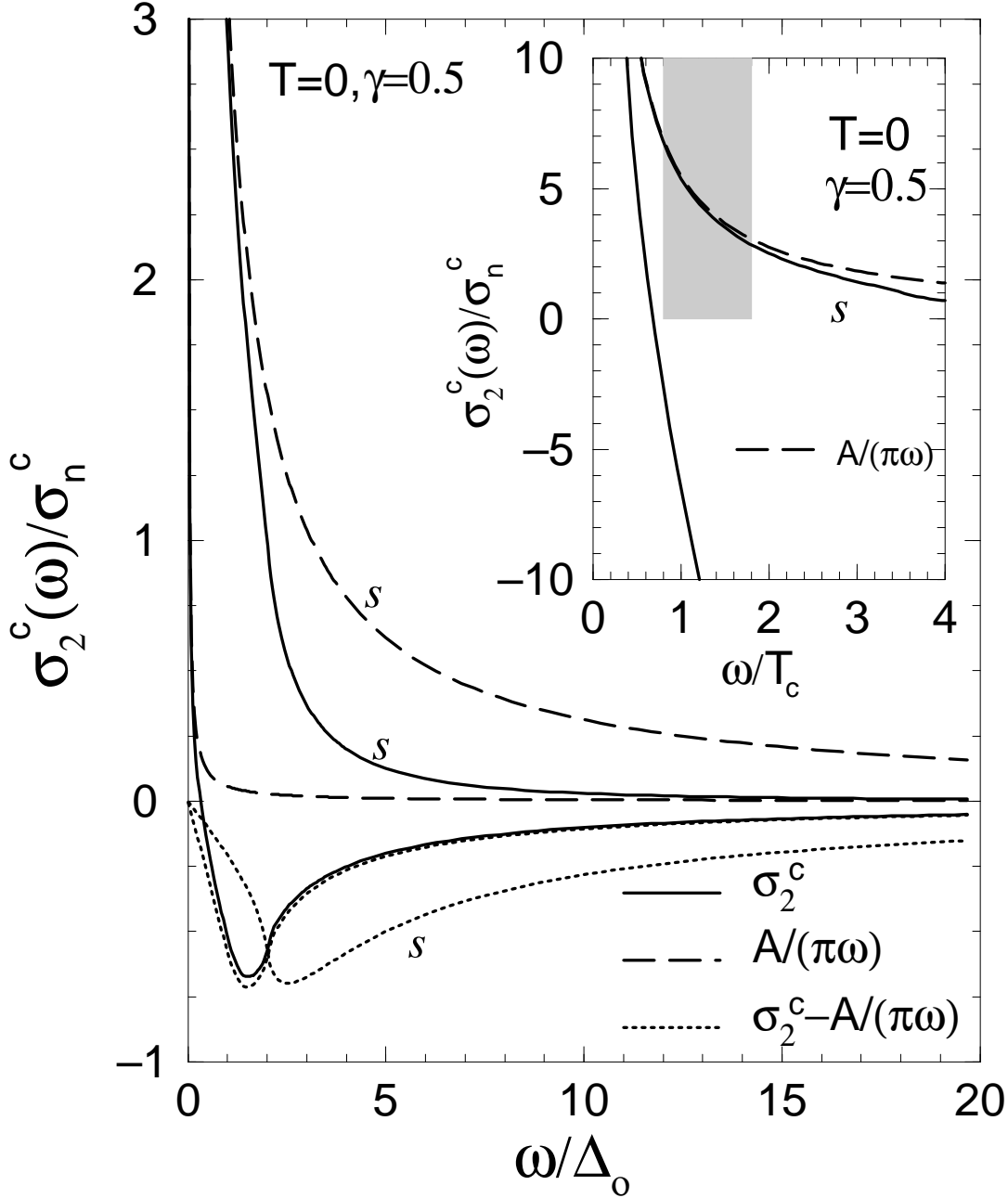}
\caption{
  The imaginary part of the $c$-axis conductivity (reactance) at $T=0$
for a scattering rate described by $\alpha=\hbar/2\pi\tau\Delta_0\to 0$.
  The $s$-wave results are denoted by an $S$.
  We show $\sigma_2^c(\omega)=\mbox{Im\,}{\sigma^\bot(\omega)}$, and separately
the London term, $A/\pi\omega$, and the difference of these two.
  The constant $A$ was extracted from $\sigma^\bot(\omega)$ by a
Kramers-Kronig transformation.
  Inset:~The $d$-wave result has been rescaled to yield the same penetration
depth, $\lambda_\bot(0)$, as for the $s$-wave result (see text).
  The shaded region indicates the range of measured plasma frequencies,
$\omega_{pc}$, in cuprate superconductors. }
\label{reactance}
\end{figure}
\end{minipage}

\end{document}